# The number, luminosity, and mass density of spiral galaxies as a function of surface brightness

**Stacy S. McGaugh**★

*Institute of Astronomy, University of Cambridge, Madingley Road, Cambridge CB3 0HA*

**ABSTRACT**

I give analytic expressions for the relative number, luminosity, and mass density of disc galaxies as a function of surface brightness. These surface brightness distributions are asymmetric, with long tails to lower surface brightnesses. This asymmetry induces systematic errors in most determinations of the galaxy luminosity function. Galaxies of low surface brightness exist in large numbers, but the additional contribution to the integrated luminosity density is modest, probably 10 – 30%.

★ Present address: Department of Terrestrial Magnetism, Carnegie Institution of Washington, 5241 Broad Branch Road NW, Washington DC 20015, USA

# 1 INTRODUCTION

The space density of Low Surface Brightness (LSB) galaxies has long been a controversial and confusing subject. It is basic to our inventory of the contents of the universe, and is crucial to many aspects of extragalactic astronomy. For example, the density of LSB galaxies has an impact on the luminosity function, faint galaxy number counts, Ly$\alpha$ absorption systems, and theories of galaxy formation. In this paper I derive analytic expressions which quantify the density of discs of all surface brightnesses.

In order to clarify some of the points of confusion, it is important to quantify what is meant by the term 'LSB'. Different people use it to mean different things, leading to contradictory interpretations of otherwise consistent data. One frequent but misleading use is treating 'LSB' and 'dwarf' as synonymous. Dwarf galaxies may be a subset of galaxies which are low in surface brightness, but there exist many LSB galaxies which can not be considered dwarfs either in terms of size or luminosity (Romanishin, Strom & Strom 1983; Phillipps & Disney 1986; Bothun et al. 1987; Davies et al. 1988; Impey & Bothun 1989; Bothun et al. 1990, Irwin et al. 1990; Knezek 1993; Sprayberry et al. 1993; McGaugh & Bothun 1994; de Blok, van der Hulst, & Bothun 1995a). Some seem to refer to LSB galaxies as things which are totally invisible optically. Others use the term to refer to any galaxy which is fainter than some canonical value, like that of Freeman (1970) or the night sky.

In order to provide a consistent definition, it is necessary to quantify the luminosity profiles of galaxies. Sérsic (1969) found that the radial luminosity distribution could be fit by

$$\Sigma(r) = \Sigma_0 e^{-(r/h)^\beta} \qquad (1)$$

where $\Sigma_0$ is the central surface luminosity density (in $L_\odot \mathrm{pc}^{-2}$), $h$ is a physical scale length (in kpc), and $\beta$ is an index which varies the shape of the profile. This is a generalization of the exponential disc ($\beta = 1$, de Vaucoulers 1959) and $r^{1/4}$ ($\beta = \frac{1}{4}$, de Vaucoulers 1948) profiles.

Since disc and bulge are believed to be distinct physical components, and most debate concerns the numbers of LSB disc galaxies, I will limit the discussion to profiles with $\beta = 1$ (i.e., spiral, irregular, and dE galaxies). It is straightforward to generalise the results presented below for different profile shapes, but in principle a multiparamter approach which adequately represents the full appearance of galaxies is required (see McGaugh, Bothun, & Schombert 1995a).

In magnitude units the exponential profile is

$$\mu(r) = \mu_0 + 1.086 \frac{r}{\alpha}, \qquad (2)$$

where $\mu_0$ is the central surface brightness and $\alpha$ is the angular scale length projected on the sky by a galaxy of physical scale length $h$ at distance $d$. The surface brightness and size of a disc galaxy are characterised by $\mu_0$ and $h$. For pure disc systems, $\mu_0$ is simply related to the effective surface brightness. I will therefore use $\mu_0$ to characterise the global surface brightness of disc galaxies.

Using $\mu_0$ it is now possible to define what is meant by low and high surface brightness (HSB). Freeman (1970) asserted that all spiral discs had the same central surface brightness,

$$\mu_0 = 21.65 \pm 0.30 \; B \text{ mag arcsec}^{-2}. \tag{3}$$

This has become widely known as 'Freeman's law', and is often used as a working definition of HSB. To be slightly more specific, I will define as high surface brightness any disc with $\mu_0 < 22 \; B$ mag arcsec$^{-2}$, which corresponds to $\Sigma_0 = 100 \; L_\odot \text{pc}^{-2}$. Any disc which satisfies this criterion is likely also to fall within the narrow range specified by Freeman's law, and the particular value is chosen to bracket it on the faint side. On the bright side, I will describe as very high surface brightness (VHSB) anything which fails to obey Freeman's law by being too bright, with $\mu_0 < 21.25$ mag arcsec$^{-2}$ or $\Sigma_0 > 200 \; L_\odot \text{pc}^{-2}$.

The most obvious dividing line for defining LSB is the brightness of the night sky, one magnitude fainter than the Freeman value in the best of conditions. To give a linear factor of two between this and HSB, and to be clearly fainter than the night sky, let us describe as LSB any disc galaxy with $\mu_0 > 22.75$ mag arcsec$^{-2}$. Fig. 1 shows examples of idealised high and low surface brightness galaxy profiles.

Though I would prefer to avoid the pollution of excessive nomenclature, further quantitative distinctions are useful. The transition region $22 < \mu_0 < 22.75$ between HSB and LSB I will call intermediate surface brightness (ISB), since objects in this range do not obey Freeman's law but are not particularly dim either. It is also useful to denote the extremes of the LSB realm by very low surface brightness (VLSB: $24.5 < \mu_0 < 27$) and extremely low surface brightness (ELSB: $\mu_0 > 27$ mag arcsec$^{-2}$). Most LSB galaxies which are known in the field are too bright for either of these categories, though some examples do exist. Nonetheless, it is possible to see VLSB galaxies on survey plates if one looks hard enough. It is not until the ELSB regime that galaxies become effectively invisible to the deepest existing photographic surveys; only one galaxy this faint has so far been reported (GP 1444; Davies, Phillipps, & Disney 1988).

The nomenclature utilised here is summarized in Table 1. While such quantitative definitions are necessary to make sensible use of terms like LSB, it should be realised that there is some *distribution* of central surface brightnesses. Obtaining this should be our goal.

## 2  THE RELATIVE NUMBERS OF LSB GALAXIES

To begin in this direction, it is possible to make a crude estimate of the relative number of galaxies in two widely separated surface brightness bins. Recent surveys (Schombert & Bothun 1988; Schombert et al. 1992) have identified large numbers of LSB galaxies by visual examination of the plates of the new Palomar Sky Survey. The selection criterion of the Schombert et al. (1992) sample (henceforth referred to as the LSB sample) mimic that of the diameter limited UGC (Nilson 1973): $\theta \geq 1'$. The only difference is the greater depth and lower grain noise of the new plates. Objects are not included if already catalogued in the UGC.

Extending the UGC in this way will incorporate galaxies which just missed the original diameter criterion. This will include HSB galaxies which are on average slightly more distant as well as lower surface brightness galaxies. In order to reject the former and isolate from the surface brightness distribution some portion which is truly LSB, only galaxies were included which were essentially undetected on the POSS-I. This guarantees that only shallow profile, LSB galaxies are selected. Surface photometry of a subset of galaxies from the LSB sample shows that this procedure selects a population which has a typical central surface brightness of $\mu_0 = 23.4$ $B$ mag arcsec$^{-2}$ (McGaugh & Bothun 1994; de Blok et al. 1995a). The distribution is very sharply peaked around this value, with a standard deviation of 0.6 and a high kurtosis.

Note that $\mu_0 = 23.4$ is $\sim 5\sigma$ fainter than the Freeman value. If Freeman's law were correct, and the surface brightness distribution really was a sharply peaked function, then these LSB galaxies should not exist. Yet they represent an $\sim 11\%$ enhancement over the number of galaxies in the UGC (Schombert & Bothun 1988), and the UGC already contains a fair number of LSB galaxies (Romanishin et al. 1983; McGaugh & Bothun 1994; de Blok et al. 1995a).

For a direct comparison to bona-fide HSB galaxies, a sample is required which has surface photometry so that the adopted definition of HSB can be quantitatively satisfied. Such a sample exists in the data of van der Kruit (1987), who claimed to confirm Freeman's law for large, early type discs. This HSB sample contains galaxies which have $\mu_0 = 21.5 \pm 0.4$ $B_J$ mag arcsec$^{-2}$, the difference between this and the Freeman (1970) value being exactly compensated by the difference between the $B$ and $B_J$ bands. There are 74 disc galaxies larger than $2'$ over 450 square degrees of sky, of which 51 were sufficiently face on to perform surface photometry. Of these, 37 qualify as Freeman discs, the remainder being classified as LSB dwarfs. Assuming that an equivalent fraction of the discs without surface photometry are HSB gives a surface density of Freeman discs of $\mathcal{N}_{\rm HSB} \approx 0.12$ per square degree.

In comparison, Schombert et al. (1992) found 198 LSB galaxies larger than $1'$ over 2375 square degrees of sky. Thirteen of these are classified as peculiar ellipticals (with shells or some other LSB extension), leaving 185 bona fide LSB galaxies. Only seven of these are dEs, the rest are spirals and irregulars. Since dEs are low surface brightness and generally have exponential profiles, I retain them but obviously this makes no difference. It is important that the sample consists mostly of spiral galaxies, as it is often stated that Freeman's law does not apply to dwarfs. This term is rarely quantified in this context, and can follow from any of morphology, size, or luminosity. The LSB sample does *not* consist of dwarfs by *any* of these definitions. The surface density of LSB spiral galaxies is $\mathcal{N}_{\rm LSB} \approx 0.08$ per square degree.

An estimate of the relative space density can be made by noting that the actual number density $n$ is related to the surface density by $n = \Omega \mathcal{N}/V$, where $\Omega$ is the solid angle and $V$ is the volume sampled by a survey. The ratio $V/\Omega$ can be estimated for these diameter limited surveys from equation (2) given values of the characteristic $\mu_0$ of each sample and the isophotal level $\mu_\ell$ at which the diameter $\theta = 2r = 1.84\alpha(\mu_\ell - \mu_0)$ is measured. Since $\alpha = h/d$, the maximum distance at which a galaxy can lie and meet the survey requirement $\theta \geq \theta_\ell$ is

$$d = 1.84 \frac{h}{\theta_\ell}(\mu_\ell - \mu_0), \qquad (4)$$

so the volume probed by the survey is

$$V = \frac{4\pi}{3}\Omega d^3. \qquad (5)$$

This suffices for the HSB survey, but not for the LSB survey because galaxies already catalogued by the UGC are specifically excluded. This places a minimum as well as maximum limit on the volume probed; to calculate this properly the volume already surveyed by the UGC must be subtracted off. In effect, the LSB survey only probes a shell of width

$$\delta = d_{\rm LSB} - d_{\rm UGC} = 1.84 \frac{h}{\theta_\ell}(\mu_\ell^{\rm LSB} - \mu_\ell^{\rm UGC}) \qquad (6)$$

stemming from the difference in the isophotal levels at which the diameters are measured. From CCD data, the diameters of the LSB survey are determined to be measured at $\mu_\ell^{\rm LSB} = 26.0 \pm 0.3$ mag arcsec$^{-2}$ (Schombert & Bothun 1988; McGaugh 1992), and those in the UGC at $\mu_\ell^{\rm UGC} = 25.3 \pm 0.7$ (Cornell et al. 1987; see also Paturel 1975; Fouqué & Paturel 1985). The volume sampled by the LSB survey is thus

$$V = 4\pi\Omega \int_{d-\delta}^{d} R^2 dR = \frac{4\pi}{3}\Omega(3d^2\delta - 3d\delta^2 + \delta^3) \qquad (7)$$

which reduces to the familiar approximation $V = 4\pi\Omega d^2 \delta$ for $\delta \ll d$.

The relative number density of LSB to HSB galaxies now follows from the observed surface densities and the volumes sampled by each survey: $n_{\rm LSB}/n_{\rm HSB} = \mathcal{N}_{\rm LSB}/\mathcal{N}_{\rm HSB} \times (V/\Omega)_{\rm HSB}/(V/\Omega)_{\rm LSB}$. Substituting the above expressions for $V/\Omega$ gives

$$\frac{n_{\rm LSB}}{n_{\rm HSB}} = \frac{\mathcal{N}_{\rm L}}{\mathcal{N}_{\rm H}} \left(\frac{\theta_\ell^{\rm L}}{\theta_\ell^{\rm H}}\right)^3 \left(\frac{h_{\rm H}}{h_{\rm L}}\right)^3 \times$$
$$\frac{(\mu_\ell^{\rm H} - \mu_0^{\rm H})^3}{\left[3(\mu_\ell^{\rm L} - \mu_0^{\rm L})^2(\mu_\ell^{\rm L} - \mu_\ell^{\rm U}) - 3(\mu_\ell^{\rm L} - \mu_0^{\rm L})(\mu_\ell^{\rm L} - \mu_\ell^{\rm U})^2 + (\mu_\ell^{\rm L} - \mu_\ell^{\rm U})^3\right]} \qquad (8)$$

where for brevity of superscripts the value specific to each catalog is denoted by L for LSB, H for HSB, and U for UGC. All of the relevant values have been discussed, except for the ratio of scale lengths $h_{\rm HSB}/h_{\rm LSB}$. The LSB catalog is dominated by objects with scale lengths similar to the HSB galaxies studied by van der Kruit (1987), so I assume $h_{\rm HSB}/h_{\rm LSB} = 1$ (Schombert et al. 1992; McGaugh & Bothun 1994; de Blok et al. 1995a).

If instead HSB galaxies are on average larger than LSB galaxies ($h_{\rm HSB}/h_{\rm LSB} > 1$), this will *increase* the relative number of LSB galaxies since $n_{\rm LSB}/n_{\rm HSB} \propto (h_{\rm HSB}/h_{\rm LSB})^3$.

Inserting the relevant values ($\mathcal{N}_{\rm LSB} = 0.08$, $\mathcal{N}_{\rm HSB} = 0.12$, $\theta_\ell^{\rm LSB} = 1$, $\theta_\ell^{\rm HSB} = 2$, $h_{\rm HSB}/h_{\rm LSB} = 1$, $\mu_0^{\rm HSB} = 21.5$, $\mu_0^{\rm LSB} = 23.4$, $\mu_\ell^{\rm HSB} = 26.5$, $\mu_\ell^{\rm LSB} = 26.0$, $\mu_\ell^{\rm UGC} = 25.3$) into equation (8) yields

$$\log\left(\frac{n_{\rm LSB}}{n_{\rm HSB}}\right) = -0.03. \qquad (9)$$

That is, there are approximately the same number of galaxies with $\mu_0 = 23.4$ as with $\mu_0 = 21.65$.

This result is obviously very different from the notion that all spiral discs have the same central surface brightness. Integrating the sharply peaked Gaussian distributions of Freeman (1970) and van der Kruit (1987) to obtain the density expected for galaxies with $\mu_0 \geq 23.4$ mag arcsec$^{-2}$ (i.e., those $\geq 5\sigma$ deviant), Freeman's law predicts

$$\log\left(\frac{n_{\rm LSB}}{n_{\rm HSB}}\right) = -6.52. \qquad (10)$$

There are several uncertainties in this calculation, but none which come close to six and a half orders of magnitude. Aside from the ratio of scale lengths, variation of the parameters in equation (8) over reasonable ranges does not cause the answer to change by more than a factor of a few (see Table 2). Even if several factors conspire together, the result is still not much affected.

The variations considered in Table 2 are rather extreme. It is very unlikely that $\mu_\ell^{UGC}$ has been misestimated by a magnitude, and it does not really matter if it has since the volume already surveyed for LSB galaxies by the UGC is essentially zero for $\mu_\ell \leq 24.3$. If, on the other hand, $\mu_\ell^{UGC}$ is even a little fainter than 25.3, the number of LSB galaxies rises sharply since $\delta \to 0$ as $\mu_\ell^{UGC} \to \mu_\ell^{LSB}$. It is not very sensible to consider variations in $\mu_\ell^{HSB}$ since all that matters to the calculation is the difference between this and $\mu_0^{HSB}$. This difference is not sensitive to zero point errors, only to improperly corrected nonlinearities. The selection isophote of the LSB sample is well determined and uniform, but it is possible that the subset of galaxies for which CCD photometry is available is not statistically representative. Even so, varying $\mu_0^{LSB}$ does not alter the answer much, and if anything it is more likely that the sample as a whole is fainter than those objects which have been studied in detail. If so, this again raises rather than lowers the LSB number density.

The ratio of typical scale lengths can make a large difference to this calculation. As already noted, it goes in the wrong sense required to save Freeman's law. There is no evidence that the distribution of scale lengths in the LSB sample of Schombert et al. (1992) is distinguishable from that in the HSB sample of van der Kruit (1987). It is thus impossible to escape the conclusion that Freeman's law is erroneous by arguing that it does not apply to 'dwarfs'.

An estimate of the error in the relative density of LSB galaxies can be made from counting statistics and Table 2. The 185 galaxies in the LSB sample contribute 0.07 dex to the error in the relative density, while the 37 bona fide Freeman discs in the HSB sample contribute 0.16 dex. This results in a combined statistical error of 0.18. From Table 2, a reasonable estimate of the errors due to fluctuations in the various $\mu_\ell$ is $\sim 0.2$ dex, and an equal amount comes from possible (small) deviations of the scale length ratio from unity. Combining all these sources of error in quadrature gives a total error of 0.34 dex. Using this number and equations (9) and (10), Freeman's law is formally rejected with a confidence of $19\sigma$. Though this error includes some estimate of the systematic as well as random uncertainties, the systematic problems could of course be larger. However, it is impossible to consider a space density of LSB galaxies less than indicated by the raw surface densities. The volume correction is fairly modest (a factor of $\sim 10$), and the LSB catalog is if anything less complete than the HSB one. Hence, a very hard lower limit is $\log(n_{LSB}/n_{HSB}) > -1$. This is still over five orders of magnitude larger than predicted by Freeman's law.

## 3  AN ILLUSTRATIVE SURFACE BRIGHTNESS DISTRIBUTION

In order to understand why studies which appear to confirm Freeman's law are so misleading, let us imagine a particular form for the surface brightness distribution and ask what we would expect to observe. As we shall see, Freeman's law arises from the failure to apply volume corrections to the observed *apparent* distribution in order to recover the intrinsic distribution (see Allen & Shu 1979 as well as Disney 1976; Disney & Phillipps 1983; McGaugh et al. 1995a).

In order to illustrate the effect surface brightness has on the volume sampled by a complete survey, let us hypothesise a simple distribution of disc galaxy central surface brightnesses, $\phi(\mu_0)$. I define $\phi(\mu_0)$ to be the space density of galaxies at each central surface brightness relative to that at some fiducial value $\mu_0^*$: $\phi(\mu_0) \equiv n(\mu_0)/n(\mu_0^*)$. Motivated by the results of §2, let us assume that galaxies exist in equal numbers faintwards of the fiducial value. Brightwards of this, let us assume a sharp exponential cut off. Hence

$$\phi(\mu_0) = \begin{cases} 1 & \text{for} \quad \mu_0 > \mu_0^* \\ e^{(\mu_0-\mu_0^*)/\gamma} & \text{for} \quad \mu_0 < \mu_0^* \end{cases} \quad (11)$$

The exponential cut off at the bright end is made to be consistent with the lack of galaxies observed with central surface brightness brighter than the Freeman value (Allen & Shu 1979). The fiducial value $\mu_0^*$ corresponds to the Freeman value. To be consistent with a narrow apparent distribution, the exponential cut off must be sharp, $\gamma \lesssim 0.3$.

In order to illustrate the effects of $\mu_0$ on diameter and magnitude selected samples below, I make two further simplifying assumptions. The first is that $\mu_0$ is not correlated with $h$, so that on average only the surface brightness matters to the selection of galaxies at each $\mu_0$. The second is that complete catalogs can indeed be constructed over sufficiently large volumes of space that they provide a fair representation of the universe containing the hypothesised $\phi(\mu_0)$.

## 3.1 Diameter Selection

For a complete, diameter limited survey, the volume correction associated with any given galaxy is obtained from $V/V_{max} = (\theta_\ell/\theta)^3$. Using equation (4), $\theta$ can be decomposed into $h$, $\mu_0$, and $\mu_\ell$, giving the volume sampled as a function of the galaxy parameters ($\mu_0, h$) and the survey parameters ($\theta_\ell, \mu_\ell$). The volume over which a particular galaxy can exist and satisfy the selection criteria follows from equation (5), $V \propto [h(\mu_\ell - \mu_0)]^3$. Larger and higher surface brightness galaxies can be detected over larger volumes, and will be numerically over-represented in catalogs relative to their space density. The effect is strong in both $\mu_0$ and $h$, though it is stronger in $\mu_0$ in the sense that no galaxy with $\mu_0 \geq \mu_\ell$ will be included in the catalog while it is in principle possible to detect galaxies with arbitrarily small $h$. Note also that $V$ only increases as $\mu_0$ becomes brighter. It never reaches a maximum at a preferred value of $\mu_0$ as suggested by Disney (1976 — see McGaugh et al. 1995a).

Specific cases are illustrated by Fig. 1. At a particular surface brightness, changing the size $h$ by a factor of two also changes the isophotal diameter by a factor of two. So the relative volume over which these galaxies could be found differs by a factor of eight, regardless of the isophotal level at which the diameters are measured (providing that this is not so faint that it is beyond the edge of the discs). At a fixed size, changing the surface brightness $\mu_0$ alters the isophotal diameter in a way that depends on the difference between $\mu_\ell$ and $\mu_0$. The brighter $\mu_\ell$, the more serious the effect. For the illustrated case of shifting $\mu_0$ by 1.5 mag arcsec$^{-2}$, the ratio of isophotal diameters are $\theta_{24}^{\text{HSB}}/\theta_{24}^{\text{LSB}} = 2.77$ and $\theta_{26}^{\text{HSB}}/\theta_{26}^{\text{LSB}} = 1.52$. Hence the relative volumes over which galaxies *of the same size* can be detected vary by a factor of 21.25 and 3.5 for $\mu_\ell = 24$ and $\mu_\ell = 26$, respectively. This illustrates the enormous sensitivity to $\mu_\ell$, and the importance of measuring it consistently and accurately when attempting to construct complete catalogs. Note also that it is essentially impossible to construct a truly complete catalog when the level $\mu_\ell$ is chosen (as is usually done) to be at the ultimate limit of the survey material, since the signal to noise for low surface brightness images will go to zero as $\mu_0 \to \mu_\ell$.

Rather than specify a particular set of survey parameters, let us focus on the relative functional effects of the galaxy profile by defining a fiducial galaxy of parameters $\mu_0^*$, $h^*$. The obvious choice for $\mu_0^*$ is the maximum in the distribution (§3.1), and $h^*$ can be chosen to make this an $L^*$ galaxy (as in Fig. 1). Normalised to this fiducial galaxy, the volume sampled by a diameter limited survey is

$$\frac{V(\mu_0, h)}{V(\mu_0^*, h^*)} = \left[\frac{h(\mu_\ell - \mu_0)}{h^*(\mu_\ell - \mu_0^*)}\right]^3. \tag{12}$$

Properly, this should be displayed in a three dimensional plot as a surface of $V$ which declines rapidly for $\mu_0 > \mu_0^*$ and $h < h^*$ (see McGaugh et al. 1995a). Since the focus here is on surface brightness, I shall restrict the discussion to simple two dimensional plots which are projections along the surface brightness axis.

The apparent number of galaxies observed at each surface brightness $N_{obs}(\mu_0)$ will just be the true distribution $\phi(\mu_0)$ convolved with the sampling function $V(\mu_0, h)$. For the intrinsic distribution described in §3.1,

$$N_{obs}(\mu_0) \propto \left[\frac{(\mu_\ell - \mu_0)}{(\mu_\ell - \mu_0^*)}\right]^3 \qquad \text{for} \qquad \mu_0 > \mu_0^* \qquad (13)$$

and

$$N_{obs}(\mu_0) \propto \left[\frac{(\mu_\ell - \mu_0)}{(\mu_\ell - \mu_0^*)}\right]^3 e^{(\mu_0 - \mu_0^*)/\gamma} \qquad \text{for} \qquad \mu_0 < \mu_0^* \qquad (14)$$

The result is shown in Fig. 2 for several values of $\mu_\ell$.

Even though the intrinsic distribution is flat faintwards of $\mu_0^*$, the apparent distribution is very strongly peaked at $\mu_0^*$, regardless of the value of $\mu_\ell$. All that is gained by pushing $\mu_\ell$ fainter is to extend the tail of the apparent distribution (cf. Allen & Shu 1979). The relative volume sampled becomes very small for $\mu_0$ two magnitudes brighter than $\mu_\ell$, so finding any galaxies at all with $\mu_0 \approx \mu_\ell - 2$ (and such objects do exist in both the UGC and the list of Schombert et al. 1992) immediately indicates a large space density of such LSB galaxies.

### 3.2 Flux Selection

For the case of galaxy selection by apparent magnitude, the survey parameters which need to be specified are the flux limit and the way in which the flux is measured. Usually, isophotal fluxes are used. By definition, these are not total fluxes, nor are they simply related to total fluxes except in the special case that the distribution of $\mu_0$ is a $\delta$-function. This has an important consequence: the flux by which a galaxy is selected is not directly related to its luminosity as with point sources, so inverting the selection function does not directly yield the luminosity function.

Nevertheless, isophotal magnitudes do have the advantage that the portion of the total flux which they represent is rigorously defined (i.e., that within the isophote $\mu_\ell$). It is harder to determine what is actually being measured by other flux measurement schemes. In principle one needs profile information to derive the total luminosity of a galaxy, and isophotal magnitudes have the virtue of being formally related to this.

The total luminosity of a galaxy can be obtained from the profile by integrating equation (1). For disc galaxies with $\beta = 1$,

$$L = \int_0^{2\pi} \int_0^{r_{max}} \Sigma_0 e^{-(r/h)} r\, dr\, d\theta. \qquad (15)$$

For $r_{max} \to \infty$ this is simply

$$L_\infty = 2\pi h^2 \Sigma_0. \qquad (16)$$

The luminosity contained within some finite number of scale lengths $x$,

$$x = r_{max}/h = 0.92(\mu_\ell - \mu_0), \tag{17}$$

is

$$L_\ell = L_\infty[1 - (1+x)e^{-x}]. \tag{18}$$

This is the true total luminosity for discs which truncate after $x$ scale lengths. Discs are observed to remain exponential for at least three scale lengths, and usually more. This encompasses most of the light ($L_\ell > 0.8 L_\infty$ for $x > 3$), so the integration to infinity is a reasonable approximation of the true total luminosity. Equation (18) can also be used to relate the flux measured within any given isophote to the total flux.

Galaxies with brighter $\mu_0$ will emit a larger fraction of their total flux above any give isophote than LSB galaxies, even of the same total luminosity. For the cases illustrated in Fig. 1, the fraction of the total flux $f = F_\ell/F_\infty$ measured by isophotal magnitudes are $F_{24}/F_\infty = 0.64$ and $F_{26}/F_\infty = 0.91$ for HSB galaxies, and $F_{24}/F_\infty = 0.18$ and $F_{26}/F_\infty = 0.74$ for LSB galaxies. Hence, LSB galaxies will have their total fluxes underestimated, and *appear* to be much fainter than HSB galaxies even if they have the *same* luminosity. For the present case, $f_{LSB}/f_{HSB} = 0.28$ at $\mu_\ell = 24$, and $f_{LSB}/f_{HSB} = 0.81$ at $\mu_\ell = 26$.

This has a serious effect on the volume probed by magnitude limited surveys. Since $V \propto L_\ell^{3/2}$, at $\mu_\ell = 24$, $V_{LSB}/V_{HSB} = 0.15$, and at $\mu_\ell = 26$, $V_{LSB}/V_{HSB} = 0.73$. Hence both the number density and the luminosity of LSB galaxies will be underestimated by magnitude limited surveys in a way which depends sensitively on the effective isophotal level at which the magnitudes are measured.

Though typical of large photographic galaxy surveys, these $\mu_\ell$ are arbitrary. So too is the choice of $\mu_0$ for LSB galaxies. Galaxies with fainter $\mu_0$ are known to exist, including the extreme case of Malin 1 with $\mu_0 > 26$ and $L > L^*$. The important point is that the identification of *any* galaxy of faint $\mu_0$ in a complete, magnitude limited sample immediately *demands* a large number density of such objects. There is nothing mysterious about this 'selection effect'; it stems simply from the fact that surveys sample a very much smaller volume of space for galaxies of low surface brightness. These galaxies will thus appear rare even if common, just as intrinsically faint stars are common even though the naked eye perceives mostly stars which are intrinsically bright. The difference is that LSB galaxies will appear rare, *even if intrinsically luminous*.

To quantify these effects for arbitrary $\mu_0$ and $\mu_\ell$, let us assume the same intrinsic density distribution as before. In general, the case of flux selection is not as clean as diameter selection, for while the diameter is dominated by the disc, the flux will also contain light from any bulge component which is not included in a simple exponential profile. However, the bulge is a small fraction of the total light for most disc systems, so the exponential profile remains a reasonable if imperfect approximation.

The volume sampled goes as $V \propto L_\ell^{3/2} = \{2\pi h^2 \Sigma_0[1-(1+x)e^{-x}]\}^{3/2}$. Normalising, as before, to a fiducial galaxy of parameters $(\mu_0^*, h^*)$, and noting that $\Sigma_0/\Sigma_0^* = 10^{-0.4(\mu_0-\mu_0^*)}$, the relative volume sampled by a flux limited survey as a function of surface brightness and size is

$$\frac{V(\mu_0, h)}{V(\mu_0^*, h^*)} = \left(\frac{h}{h^*}\right)^3 10^{-0.6(\mu_0 - \mu_0^*)} \left[\frac{1 - (1+x)e^{-x}}{1 - (1+x^*)e^{-x^*}}\right]^{3/2}. \qquad (19)$$

For the assumed intrinsic distribution, the observed distribution $N_{obs}(\mu_0) = \phi(\mu_0)V(\mu_0)$ is shown in Fig. 3.

The apparent distribution for flux selected catalogs is even sharper than for the diameter limited case. This is because of the exponential rather than power law dependence on $\mu_0$. Unlike diameter limited catalogs, where the difference between $\mu_\ell$ and $\mu_0$ is what matters, it is the difference between $\mu_0$ and $\mu_0^*$ that matters most to flux selection. Galaxies with low surface brightnesses will be very strongly selected against because of the faintness of $\mu_0$, *regardless* of $\mu_\ell$. This only enters through $x$, and matters little. Thus, diameter limited surveys in principle provide a much better picture of the true galaxy population.

## 4 PREVIOUS RESULTS

With the results of §3, it is possible to understand why the issue of the number density of LSB galaxies, and the surface brightness distribution in general, has remained confused for so long. This results largely from unquantified use of the term LSB, and from the misuse, or total lack, of volume corrections. In this section, I review earlier work and show that essentially all data sets provide consistent results when properly analysed.

### 4.1 Optical Studies

Discussion of the role of surface brightness in galaxy selection goes back at least to Zwicky (1957), Arp (1965), and de Vaucoulers (1974). The first quantitative statement about the distribution of galaxy surface brightnesses was in effect made by Freeman (1970). Complete samples did not exist at the time, so no correction for volume sampling was considered.

Many people, especially Disney (1976), Disney & Phillipps (1983), Phillipps et al. (1987) and Davies (1990), pointed out that selection effects could cause Freeman's law. However, these arguments were based on the assumption of no correlation between luminosity and surface brightness. While galaxies cover a wide range in the $\mu_0$–$L$ plane, current data show that lower surface brightness galaxies do tend to be less luminous, just not smaller. In this case the qualitative effects are very different (see McGaugh et al. 1995a for a detailed comparison). The same effect had already been noted by Allen & Shu (1979) who pointed out that Freeman's law is an overinterpretation of the data: all that can really be said is that galaxies brighter than the Freeman value are rare; the number fainter remained unconstrained.

Freeman's law was apparently confirmed by many workers (Freeman 1970; Schweizer 1976; Thuan & Seitzer 1979; Boroson 1981; van der Kruit 1987), but only van der Kruit (1987) claimed to have a complete sample with surface photometry and further claimed that Freeman's law only applied to large, early type galaxies. He found a lack of galaxies with $\mu_0 \gtrsim 22.5\ B_J$ mag arcsec$^{-2}$ and $h \gtrsim 2$ kpc (for $H_0 = 100\,\mathrm{km\,s^{-1}Mpc^{-1}}$). The galaxies which were LSB were also small and late (Sd or later) type, opposite the results of Kent (1985). Nonetheless, van der Kruit (1987) concluded that selection effects *as formulated* by Disney & Phillipps (1983) were not occurring. Phillipps et al. (1987) state that van der Kruit (1987) misapplies their formalism, a point pursued by Davies et al. (1994). The latter claim to be able to reproduce the apparent distribution of the data of van der Kruit (1987) with the Disney & Phillipps (1983) visibility formalism if there is an implicit magnitude limit in his diameter limited catalog because of signal to noise constraints. However, if this reasoning were correct, van der Kruit (1987) would not have found the LSB dwarfs which he did.

There is no obvious reason why van der Kruit (1987) missed large LSB galaxies. His study is unique in this point (Romanishin et al. 1983; Davies et al. 1988; Irwin et al. 1990; McGaugh & Bothun 1994; de Blok et al. 1995a; de Jong & van der Kruit 1994). If anything, the largest scale length galaxies tend to be low rather than high surface brightness (Bothun et al. 1990; Knezek 1993; Sprayberry et al. 1995a), the only consensus point being that the lack of large, HSB disks is *not* a selection effect (de Jong & van der Kruit 1994). The large diameter limit employed by van der Kruit (1987) can only survey a very limited volume of space for galaxies dimmer than the Freeman value, so perhaps this region does not constitute an adequate sample of the universe.

Though the statistics are poor, the apparent distribution of central surface brightnesses in the data of van der Kruit (1987) is roughly what is expected from Fig. 2. There is one large LSB galaxy in his sample (NGC 4392) which, when corrected for volume sampling, implies a greater density than all the large Freeman discs combined (see his Fig. 6). Since there was only one such galaxy in his sample, van der Kruit (1987) chose to ignore it. Obviously, this invalidates his primary conclusion confirming Freeman's law, even when restricted to large discs.

Aside from the possibility of Freeman's law being a real effect or due to selection effects, it has also been suggested that it could be caused by improper disc-bulge deconvolution (Kormendy 1977; Phillipps & Disney 1983; Schombert & Bothun 1987; Davies 1990; Rönnback 1993) or inclination and internal extinction (e.g., Peletier & Willner 1992; see also Giovanelli et al. 1995). While important, neither of these effects are really pertinent here. They are usually small, and almost never as large as the 2 mag. difference between Freeman's law and the LSB galaxies of Schombert et al. (1992). Moreover, they do not matter much to selection, which is by far the dominant effect.

More recently, Bosma & Freeman (1993) and Roukema & Peterson (1995) have addressed this problem directly with further surveys. The results of Bosma & Freeman (1993) are examined in detail below. Roukema & Peterson (1995) claim to find a small number ($\sim 6\%$) of LSB galaxies. However, they fail to quantitatively define LSB, or to isolate a population which could be described as such (though their Fig. 5 does illustrate the absurdity of Freeman's law). Their selection criteria are subjective, as is their estimate of the redshift of completeness. This has a strong influence on the results since $V \propto z_c^3$ — a no less arbitrary choice of $z_c$ would give a number of 100%. In essence, they make both major mistakes which have befuddled this field: no quantitative definition of terms, and improper correction for volume sampling. When these problems are considered, essentially *all* published data are consistent despite the many contradictory and misleading interpretations given them.

## 4.2 The Diameter Distribution

Bosma & Freeman (1993) approach the problem of the central surface brightness distribution by examining the distribution of diameter ratios on plates of different depth. They measure only the apparent distribution, without correction for the volume sampled. In this section, I derive this correction and apply it to their results.

The ratio of the diameter of the same galaxy measured on two plates of different limiting isophotes $\mu_{\ell_1}$ and $\mu_{\ell_2}$ is simply related to the central surface brightness $\mu_0$. In effect, this is a very crude form a surface photometry which fits a straight line to two points to estimate $\mu_0$. Bosma & Freeman (1993) define the parameter $\Gamma$ to be the observed diameter ratio, which is related to $\mu_0$ by

$$\Gamma = \frac{\theta_{\ell_1}}{\theta_{\ell_2}} = \frac{\mu_{\ell_1} - \mu_0}{\mu_{\ell_2} - \mu_0}. \tag{20}$$

The samples are selected by diameter on the deeper plates (by choice $\mu_{\ell_1} > \mu_{\ell_2}$). Bosma & Freeman (1993) employ plates from the first Palomar sky survey ('Pal'), the ESO-B survey ('ESO') and SRC-J survey ('SRC'). They derive values for the limiting isophotes of these materials of $\mu_\ell^{Pal} = 24.62$, $\mu_\ell^{ESO} = 25.10$, and $\mu_\ell^{SRC} = 25.59$ essentially by demanding that Freeman's law be true. This causes their derived values to differ significantly from the actual measured values: $\mu_\ell^{Pal} = 25.3$ (Cornell et al. 1987), $\mu_\ell^{ESO} \approx 26$ (Lauberts & Valentijn 1989), and $\mu_\ell^{SRC} \approx 27$ (Corwin, de Vaucoulers, & de Vaucoulers 1980). However, the numerical value of $\mu_\ell$ is not as important as their failure to consider what one should expect to see.

The expected apparent distribution for the intrinsic model of §3.1 can be derived by substituting $\Gamma$ for $\mu_0$ in equations (13) and (14). From the definition of $\Gamma$,

$$\mu_0 = \frac{\Gamma \mu_{\ell_2} - \mu_{\ell_1}}{\Gamma - 1}. \tag{21}$$

For a flat intrinsic surface brightness distribution, the apparent distribution will be

$$N_{obs}(\Gamma) = \left( \frac{\mu_{\ell_1} - \frac{\Gamma \mu_{\ell_2} - \mu_{\ell_1}}{\Gamma - 1}}{\mu_{\ell_1} - \frac{\Gamma^* \mu_{\ell_2} - \mu_{\ell_1}}{\Gamma^* - 1}} \right)^3, \tag{22}$$

where $\Gamma^*$ corresponds to $\mu_0^*$. This is plotted in Fig. 4 together with the data of Bosma & Freeman (1993).

Even for a flat intrinsic distribution, the expected apparent distribution $N_{obs}(\Gamma)$ is very strongly peaked at $\Gamma^*$. Indeed, the expected distribution is so sharp that it is necessary to consider the spread introduced by random errors in $\Gamma$ (without which values of $\Gamma < 1$ are impossible). This is complicated, because $\Gamma$ is not directly measured, but rather is the ratio of two diameter measurements. This results in a dependence of $\sigma_\Gamma$ on $\Gamma$ and a non-Gaussian error distribution. This, combined with the small expected tail towards large $\Gamma$ causes a fair number of points to be scattered to high $\Gamma \sim 1.5$. A reasonable fit to the data is obtained for $\sigma_\Gamma \approx 0.08$, which is slightly smaller than the random error estimate ($\sigma_\Gamma \approx 0.12$) of Bosma & Freeman (1993). If the errors were this large, a greater number of galaxies would be scattered to $\Gamma < 1$ than actually are.

From the apparent distribution, Bosma & Freeman (1993) conclude that 80% of spirals obey Freeman's law. They also claim that a quarter of galaxies possess discs which truncate after only two scale lengths in order to reconcile the observed spike with a Freeman law. These are both erroneous. In fact, the observations have precisely the form which is expected for an intrinsicly flat distribution. Their observation of some points with $\Gamma > 2$ suggest a high density of VLSB galaxies. The identification of a galaxy similar to Malin 1 in their complete sample also implies a large density of such galaxies.

However, $\Gamma$ provides only a very crude measure of $\mu_0$, with errors dominating the entire shape of the distribution. Indeed, the sharp spike does not even provide a good estimate of $\mu_0^*$, since the error in this goes as $(\Gamma^* - 1)^{-1}$. Since $\Gamma^*$ is always near one, small uncertainties in the position of $\Gamma^*$ lead to big ones in $\mu_0^*$. Thus all that can really be said from these data is that they are consistent with an approximately flat distribution. Any galaxies with $\Gamma > 1.3$ *rule out* a Freeman law.

### 4.3 Constraints from 21 cm Surveys

An important constraint on the population of galaxies which are difficult to detect optically is provided by 21 cm surveys. These depend on the H I rather than optical properties of galaxies, so in principle suffer no bias against gas rich but low surface brightness galaxies. However, sensitivity limits have restricted these surveys to very small volumes.

The 21 cm surveys fall into two categories. The first type are serendipitous detections of signal in the off beams of pointed observations of optically known galaxies (Fisher & Tully 1981; Briggs 1990). These provide the apparently most impressive constraints on gas rich but optically faint galaxies. The second type are blind surveys, largely independent of known optical galaxies (Kerr & Henning 1987; Weinberg et al. 1991; Szomoru et al. 1995). The latter are in principle more useful, but are to date very limited in extent.

Strictly speaking, the first type of survey only constrains the mass density of 'intergalactic H I clouds' (Fisher & Tully 1981), i.e., objects which are totally invisible optically. If a galaxy is visible at all at the serendipitous 21 cm position, regardless of its actual surface brightness, it does not count towards this invisible population. The eye is very good at detecting extended coherent structure if told where to look, so the requirement of invisibility is quite restrictive, probably $\mu_0 > \mu_\ell$.

The diameters listed in the UGC are typically measured at the $\mu_\ell \approx 25.3$ mag arcsec$^{-2}$ level, but are sometimes as faint as 27 mag arcsec$^{-2}$ (Cornell et al. 1987). Hence quite LSB galaxies are at least discernible on the POSS-I survey prints, even if their appearance thereon may seriously underestimate their true size. An object must be well into the ELSB regime to guarantee evading optical detection entirely. Thus the first set of surveys provide no real constraint at all on the density of LSB galaxies. However, data of this sort *would* be very useful if the associated optical identification criteria were quantified, and the surface brightnesses of detected galaxies actually measured.

The case of the large H I cloud in Virgo (H I 1225+0146) is a good example of the problems involved in using these surveys to place limits on the LSB galaxy population. Originally thought to be the first discovery of an H I cloud totally devoid of an optical counterpart (Giovanelli & Haynes 1989), H I 1225+0146 does have an associated optical component (McMahon et al. 1990; Impey et al. 1990). Though the H I properties of this galaxy are extreme, the optical component is actually quite prominent by the standards of Schombert et al. (1992), having a diameter $\theta_{26} \approx 2'$, (Salzer et al. 1991). This is twice the limit of Schombert et al. (1992), and H I 1225+0146 would have been incorporated as one of the more conspicuous LSB galaxies had that survey extended far enough south. Clearly, claims of strict limits on the LSB galaxy population based on these sort of data are meaningless without rigorous quantification of both terminology and optical search limits.

Blind surveys are in principle much more useful, but to date cover only very small volumes. This causes large uncertainties in normalization which make interpretation of their results difficult (see Schade & Ferguson 1994; Ferguson & McGaugh 1995). In particular, these surveys fail to recover the H I mass function which is known to exist in optically bright galaxies. Hence their ability to constrain optically dim galaxies is suspect at best. (Contrast the last two sentences of the abstract of Weinberg et al. 1991).

Szomoru et al. (1995) present more complete results from the same survey as Weinberg et al. (1991), including measurements of the optical surface brightnesses of 11 of the H I selected galaxies. Of these, three are LSB. This is a fairly high fraction, especially considering that the fields were not entirely randomly selected. Those centered on known bright galaxies are unlikely to detect LSB galaxies, which tend to be very isolated on the relevant scales (Bothun et al. 1993; Mo et al. 1994). Two of the three LSB galaxies were discovered in blank fields.

Obviously, not a great deal can be said with such limited statistics. Nonetheless, the conclusion of Szomoru et al. (1995) are consistent with those presented here. Number density and mass density are not the same thing; LSB galaxies could be quite numerous and still contain only a modest amount of mass. Moreover, as noted above, the vast majority of LSB galaxies should be visible on sky survey prints. Thus they do not represent some kind of mysterious, entirely new population, but rather a familiar if neglected one consisting predominantly of late morphological types which span and extend a continuum of galaxy properties.

## 5 THE DISTRIBUTION FUNCTIONS

In this section, I derive expressions for the relative number, luminosity, and mass density of disc galaxies as a function of central surface brightness.

### 5.1 The Number Density

The number density of galaxies can be extracted from the apparent distribution of central surface brightnesses in a complete survey by inverting the procedure described in §3. That is, given $N_{obs}(\mu_0)$, one can obtain the relative number density distribution from $\phi(\mu_0) = N_{obs}(\mu_0)/V(\mu_0)$. Unfortunately, very little data exist which meet both requirements of completeness and surface photometry, particularly for local samples of field galaxies.

One study that does is Davies (1990). These data behave precisely as expected from Figs. 2 and 3 (see McGaugh et al. 1995a). The distribution $\phi(\mu_0)$ obtained from these data is shown in Fig. 5.

The distribution on either side of the peak is well fit by a straight line. It is therefore possible to write an analytic expression that gives the relative number density of disc galaxies as a function of surface brightness:

$$\log[\phi(\mu_0)] = m(\mu_0 - \mu_0^*). \tag{23}$$

The peak is at $\mu_0^* = 21.9 B_J$ mag arcsec$^{-2}$, and by least squares fit giving equal weight to diameter and flux selected samples, the slope is

$$m = \begin{cases} -0.3 & \text{for} \quad \mu_0 > \mu_0^* \\ 2.6 & \text{for} \quad \mu_0 < \mu_0^* \end{cases} \tag{24}$$

Formally, the errors on the fit are small, $\pm 0.07$ for the slope on the faint end and $\pm 0.2$ for the bright end. However, systematic effects plus the relatively small numbers of galaxies in the fainter bins makes the actual uncertainty rather larger.

The number density derived from the data of Schombert et al. (1992) in §2 suggest a faint end slope closer to $m = 0$. There are nearly as many galaxies in this bin at $\mu_0 = 23.4$ as in the entire sample of Davies (1990), so $m = 0$ is at least as significant.

### 5.1.1 Correlations Between Surface Brightness and Size

The systematic effect of greatest concern is that the volume sampling function depends on size as well as surface brightness. So far, I have assumed that size and surface brightness are uncorrelated, so that on average variations in $h$ cancel out and $V(\mu_0, h)$ can be approximated by $V(\mu_0)$. This assumption is the natural one to make given the functional separation of the volume sampling into size and surface brightness terms. More importantly, it is motivated by the data, which show no indication of a correlation between these two parameters (Romanishin, Strom & Strom 1983; Davies et al. 1988; Irwin et al. 1990; McGaugh & Bothun 1994; de Blok et al. 1995a; McGaugh, Schombert, & Bothun 1995b). However, a weak trend one way or the other is not ruled out, so it is important to examine the consequences of such a relationship.

The volume sampling depends on scale length as $V(h) = h^3$ for both diameter and flux selection. Small galaxies are hard to find just as dim ones are. Strictly speaking, the slope $m$ describes the projection of the bivariate distribution $\Phi(\mu_0, h)$ along the surface brightness axis. To account for possible correlations with size, we should write

$$\log[\phi(\mu_0)] = m(\mu_0 - \mu_0^*) - 3\log\left(\frac{h}{h^*}\right). \tag{25}$$

If size and surface brightness *are* correlated in the sense suggested by much unwritten common lore, namely, that 'dwarfs' are both smaller *and* lower in surface brightness than brighter 'giant' galaxies, then both size and surface brightness discriminate against them. Correcting for this additional factor due to $h$ would require *even more* low surface brightness galaxies, raising the distribution in Fig. 5 and causing the fit $m$ to be less negative.

On the other hand, if size and surface brightness are correlated in the sense that lower surface brightness galaxies tend to be larger, then their increased size tends to offset their faint surface brightness. This would push the distribution in Fig. 5 downwards, towards more negative values of $m$. However, this would imply that the lowest surface brightness galaxies are quite large and luminous. Local surveys with relatively bright limiting isophotes would give a very misleading impression of this population.

There is some evidence pointing in both directions. The largest disc galaxies are inevitably those lowest in surface brightness (Impey & Bothun 1989; Knezek 1993; Sprayberry et al. 1995a) so that disc galaxies exist all over the surface brightness–luminosity diagram up to maximums in both (McGaugh 1995). However, there may be a slight tendency for the scale length distribution to become steeper with fainter surface brightness. This would boost both the number of low surface brightness galaxies in Fig. 5 and steepen the faint end slope of the luminosity function (see Bothun et al. 1991; McGaugh 1994). There is not a sufficiently strong effect in either direction to alter the conclusions reached here, though obviously these effects will modulate the precise value of $m$ somewhat.

### 5.1.2 *Cosmological Dimming*

Another systematic effect of concern for the Davies (1990) sample is cosmological dimming. The magnitude selected sample is limited at $B_J \leq 19.1$, where the median redshift is $z \sim 0.1$. At this redshift, $(1+z)^4$ dimming amounts to $\sim 0.4$ mag., plus the $k$-correction appropriate to each galaxy. Hence, it is possible that the galaxies in the faintest surface brightness bins are actually higher surface brightness galaxies which have been dimmed to appear low surface brightness. In this case, it is obviously inappropriate to apply a large volume correction to derive the number of galaxies in that low surface brightness bin.

In order to test the potential effects on the Davies (1990) sample, I have taken the empirical redshift distribution for the observed magnitude range (Koo & Kron 1992) and attributed the maximum amount of dimming to each galaxy. Starting in the faintest bins and working brightwards, each galaxy is corrected by the maximum amount allowable for the observed redshift distribution. This is not a statistical procedure; it is the worst case scenario for the analysis presented here. The results of this exercise are shown in Fig. 6.

If maximum credit is given to cosmological dimming in this fashion, something resembling a Freeman law plus a tail to lower surface brightness can be regained. However, this is a by choice a pathological case, the most significant aspect of which is that it is still impossible to recover a true Freeman law. Indeed, a universe full of Freeman discs would not dim to have the observed apparent distributions (McGaugh et al. 1995a). Moreover, surface brightness and redshift are not observed to be correlated in any published sample, and the distribution derived from this contrived exercise is inconsistent with the results of

§2. The sample of Schombert et al. (1992) gives a high density of LSB galaxies which are known to be local with negligible cosmological dimming.

Indeed, one might expect the opposite effect to be at least as important. That is, from the nature of the volume sampling function, the bona-fide LSB galaxies which are found in any complete sample should on average be of lower redshift than higher surface brightness galaxies simply because the latter can be seen further away. Repeating the above procedure in reverse, this time attributing the highest redshifts to the highest surface brightness galaxies, we find that the distribution is not so strongly affected. This is because the bulk of the observed galaxies are clustered around the same surface brightness ($\mu_0^*$), and most of the redshift distribution is filled before the low surface brightness bins are reached. This stretches the surface brightness distribution, causing it to become approximately flat. This would bring it into agreement with the density derived from the data of Schombert et al. (1992), so if anything, the latter case is likely to be more important. However, neither of these extreme cases are likely, and mostly dimming effects probably just shuffle the high surface brightness bins around a bit.

The most significant effect of dimming is not the redistribution of the shape of the surface brightness distribution so much as the shift caused by the median amount of dimming. This is $\gtrsim 0.4$ mag., depending on the colours of the galaxies. This is the amount needed to bring the value of $\mu_0^*$ determined from the Davies (1990) observations into agreement with local samples (van der Kruit 1987).

Hence, the surface brightness distribution $\phi(\mu_0)$ is well described by two straight lines which meet at $\mu_0^* = 21.5\ B_J$ mag arcsec$^{-2}$. The slope of the faint end is probably in the range $-0.3 < m < 0$ with the latter value perhaps somewhat more likely. The slope is much steeper on the bright end, with $m \approx 2.6$. (Note that this corresponds to $\gamma = 0.17$, sharper even than assumed in §3.) Significant uncertainties remain in all of these parameters, and even $\mu_0^*$ is not tremendously well determined.

### 5.1.3 Recent Data

Since the initial submission and presentation of these results (McGaugh 1993), much work has been done which strongly confirms them. Schwartzenberg et al. (1995) find a distribution which is flat until a sudden dramatic rise in the numbers of VLSB galaxies. Their data are very deep images obtained with the AAT f/1 camera, and the result is sensitive to dimming effects. Dalcanton (1995) identified a large number of VLSB galaxies in deep drift scans consistent with a flat distribution, but is obliged to make the same assumption about the distribution of scale lengths as I have done. Though not quite as deep, there are two studies (Sprayberry 1994; de Jong 1995) which do have redshifts and hence need make *no assumptions*. Both these data sets are consistent with the distribution derived here, though perceptible differences exist. The data of de Jong (1995) include extensive CCD imaging of a sample selected from the UGC and indicate a slope very similar to that of the Davies (1990) data. The data of Sprayberry (1994) are selected from plate scans by the APM, and indicate something closer to $m = 0.1$. Part of the difference may be attributed to morphological selection: de Jong (1995) concentrates on spiral galaxies, as classified by Nilson (1973). It is my qualitative impression (McGaugh et al. 1995b) that spiral structure tends to drop in amplitude (or at least visibility) with surface brightness so that Sprayberry (1994) may simply be including more smooth LSB disks. The data of both Sprayberry (1994) and de Jong (1995) indicate a somewhat brighter $\mu_0^*$ and steeper cut off than found here, but these parameters are correlated and depend on the binning and photometric system. Given the current state of the data, all results appear to be consistent.

## 5.2 The Luminosity Density

Once the number density of disc galaxies as a function of surface brightness is known, it is straightforward to obtain the luminosity density contributed by galaxies of each central surface brightness. Let us define the relative luminosity density $J(\mu_0)$ in analogy with the surface brightness distribution $\phi(\mu_0)$ so that it is normalised to the absolute luminosity density $j^*$ produced by $\mu_0^*$ galaxies. That is, $J(\mu_0) = j(\mu_0)/j(\mu_0^*)$. If the number density of galaxies at each $\mu_0$ is known, $J$ is simply the product of the number times the relative luminosity of each galaxy $\mathcal{L}(\mu_0, h)$, i.e., $J(\mu_0) = \phi(\mu_0)\mathcal{L}(\mu_0, h)$. The relative luminosity follows from equation (16), giving

$$\mathcal{L}(\Sigma_0, h) = \frac{\Sigma_0}{\Sigma_0^*} \left(\frac{h}{h^*}\right)^2 \tag{26}$$

and

$$\log[\mathcal{L}(\mu_0, h)] = -0.4(\mu_0 - \mu_0^*) + 2\log\left(\frac{h}{h^*}\right). \tag{27}$$

Combining equations (25) and (27), the relative luminosity density is

$$\log[J(\mu_0)] = (m - 0.4)(\mu_0 - \mu_0^*) - \log\left(\frac{h}{h^*}\right). \tag{28}$$

This is plotted in Fig. 7 for $h/h^* = 1$.

Basically, this just says that even if the *number* density of LSB galaxies is large ($m = 0$), the *luminosity* density produced by progressively dimmer discs declines with their surface brightness. Note that $J(\mu_0)$ does not depend strongly on the assumption about $h$ since the terms that account for the number ($\phi \propto h^{-3}$) and luminosity ($\mathcal{L} \propto h^2$) nearly offset. The numerical effect is stronger, so if lower surface brightness galaxies are on average smaller, they contribute even more to the luminosity density.

Once $J(\mu_0)$ is known, it is simply a matter of the definition of terms to determine how much luminosity density is produced by low surface brightness galaxies. If LSB is taken to mean anything fainter than the Freeman value, then well over half the luminosity comes from low surface brightness galaxies. For the opposite extreme definition of essential invisibility (ELSB), almost no light is produced ($< 1\%$ by extrapolation of the trend in Fig. 7).

Obviously, very different interpretations will follow from the same data if the terminology is not quantitatively defined. By the definitions adopted in Table 1, most of the luminosity density is produced by HSB and ISB discs, while very little is produced by either VLSB or VHSB discs. A small, but significant amount (10 − 30%) is produced by truly LSB galaxies.

### 5.3 The Mass Density

The relative mass density $\rho(\mu_0)$ follows from the product of the luminosity density and the mass to light ratio $\Upsilon$. Typically, this is assumed to be the same for all galaxies. However, one can do better with information about the systematics of the rotation properties of galaxies with surface brightness (Sprayberry et al. 1995b; Zwaan et al. 1995; de Blok, McGaugh, & van der Hulst 1995b; McGaugh et al. 1995c; Salpeter & Hoffman 1995). The observations require

$$\Upsilon \propto \Sigma_0^{-q/2} \propto 10^{0.2q(\mu_0-\mu_0^*)} \qquad (29)$$

where $q$ is a bandpass dependent parameter of order unity. Given this relation between the mass to light ratio and the surface brightness, $\rho = \Upsilon J$ is simply

$$\log[\rho(\mu_0)] = \log[J(\mu_0)] + 0.2q(\mu_0 - \mu_0^*). \qquad (30)$$

Substituting the above results as before yields

$$\log[\rho(\mu_0)] = [m - 0.2(2 - q)](\mu_0 - \mu_0^*) - \log\left(\frac{h}{h^*}\right) \qquad (31)$$

which is plotted in Fig. 8 for $q = 1$ and $h/h^* = 1$.

The mass density contained in LSB galaxies is an even greater proportion of the total than is the luminosity density. Note, however, that the observed $\Upsilon$-$\Sigma_0$ relation applies to the mass enclosed within the edges of the discs. Extrapolating this to the total mass of the halos associated with the galaxies would require knowledge of the total extent of the halos relative to the discs. The obvious assumption is that there is no systematic dependence of the halo extent on surface brightness (i.e., $R_{halo} \propto R_{disc}$).

The relatively large mass fraction in the halos of LSB galaxies suggested by Fig. 8 is consistent with the predictions of the structure formation model developed by Mo et al. (1994) to explain the spatial distribution of LSB galaxies (see their Fig. 11) and with the expectations of galaxy formation theory generally (Frenk et al. 1987; McGaugh 1992; Bothun et al. 1993; Antonuccio-Delogu 1995; Dalcanton et al. 1995).

### 5.4 The Gas Density

In addition to the relation of enclosed dynamical mass to surface brightness, there exists a very similar relation for neutral gas mass fraction (de Blok et al. 1995b). Lower surface brightness galaxies are progressively more gas rich. This may indicate an evolutionary sequence (McGaugh 1995), with galaxies increasing in surface brightness as they convert gas into stars. The scatter in the $(M_{HI}/L)$-$\Sigma_0$ relation is rather larger than that in $\Upsilon$-$\Sigma_0$, and it is possible that LSB galaxies with low gas content have been selected against.

Nevertheless, the $(M_{HI}/L)$-$\Sigma_0$ relation represents at least the upper envelope of the space occupied by galaxies. In analogy with equation (31),

$$\log[\rho_{gas}(\mu_0)] = [m - 0.2(2 - q_{gas})](\mu_0 - \mu_0^*) - \log\left(\frac{h}{h^*}\right). \tag{32}$$

If there remain many undetected gas poor LSB galaxies, $q_{gas} < q$ and $q_{gas}$ could conceivably be negative. However, from present data they are only marginally distinguishable (de Blok et al. 1995b) with $q_{gas} \approx q \approx 1$, so Fig. 8 suffices to display $\rho_{gas}(\mu_0)$ as well as $\rho(\mu_0)$.

## 6 DISCUSSION

The relationships set out in §5 quantify an important, previously unappreciated aspect of the general field galaxy population. These relations are as basic to understanding galaxies as the luminosity and correlation functions. As such, they provide an important new constraint on theories of galaxy formation and evolution. They are also relevant to various topical problems discussed below.

### 6.1 The Luminosity Function

It has long been recognized (Zwicky 1957; de Vaucoulers 1974; Disney 1976) that LSB galaxies can have an impact on determinations of the galaxy luminosity function. Generally, these have implicitly treated the surface brightness distribution as a $\delta$-function (see McGaugh 1994; Ferguson & McGaugh 1995). This is only adequate if the surface brightness distribution is symmetric, so it is interesting to see what effects the actual $\phi(\mu_0)$ may have.

Galaxies of the same luminosity but different surface brightness are sampled over different volumes (recall Fig. 1). This means that the luminosity function can not be obtained directly from the inversion of the selection function because the selection magnitude is not simply related to the total magnitude. Since all surveys have an effective limiting isophote, direct construction of $\Phi(M)$ from them always begs the question of how much light is produced by galaxies with $\mu_0 > \mu_\ell$. However, missing LSB galaxies entirely turns out not to be the most serious effect. Most surveys are able to at least detect LSB and even VLSB galaxies, and there is little luminosity density beyond that (Fig. 7). The more important effect comes from the systematic underestimation of the fluxes of LSB galaxies (this happens *by definition* with isophotal magnitudes — see equation 18). This effect can be large ($\sim 2$ mag., McGaugh & Bothun 1994) depending on $\mu_0$ and $\mu_\ell$ (McGaugh 1994). One expects LSB galaxies to compose a very small fraction of the total sample (Figs. 2 and 3), so this effect can easily go unnoticed. It is quite serious though, because the few detected LSB galaxies carry a volume normalised weight comparable to the entire rest of the survey. Small errors in their luminosity are thus magnified enormously when estimating $\Phi(M)$. This wreaks havoc on the derived Schechter (1976) parameters (McGaugh 1994; Ferguson & McGaugh 1995). The real situation may be even worse than described in these papers, because we assumed that everything had been done correctly to a uniform isophotal selection level. In practice this is not the case, as $\mu_\ell$ fluctuates from plate to plate over a range that causes significant differences in the fraction of the total flux actually measured. The effects of varying isophotal selection levels can clearly be seen in the combined data of Ellis et al. (1995). Surveys with brighter limiting isophotes give flatter slopes $\alpha$ and lower normalizations $\phi^*$ than do deeper surveys. How much of this is caused by surface brightness dependent measurement effects, large scale structure, and evolution is difficult to say; probably all contribute.

It is thus difficult to estimate the degree to which the integrated luminosity density is underestimated by current surveys. The required information is not usually published, nor even extracted and retained in the original work. Fairly complete information exists for the APM survey (Maddox et al. 1990a), for which $24.5 < \mu_\ell < 25$. For these measurement isophotes the flux will begin to be significantly underestimated in the ISB regime. From Fig. 7, I estimate that this is likely to lead to an integrated luminosity density which is underestimated by approximately 30%. This procedure is highly uncertain, and the underestimate could be as much as a factor of two (Sprayberry 1994; Dalcanton 1995), or as little as 10%. It is certainly not an order of magnitude or zero.

In principle, one should construct the bivariate distribution $\Phi(\mu_0, h)$ and integrate over this more general quantity (see Sodré & Lahav 1993; de Jong 1995). The slope of the faint end of the luminosity function is particularly sensitive to this procedure. For a flat $\phi(\mu_0)$, a flat ($\alpha = -1.0$) luminosity function (e.g., Loveday et al. 1992) requires a size distribution $\phi(h) \propto h^n$ with $n = -1.35$ (see Bothun et al. 1991). Most estimates of the size distribution (van der Kruit 1987; Hudson & Lynden-Bell 1991; de Jong 1995) are closer to $n = -2$ which corresponds to $\alpha = -1.5$, but considerable uncertainty exists. Moreover, there is little reason to expect plausible forms of the bivariate distribution to integrate to precisely the Schechter (1976) form, and while this might be a tolerable approximation, more complicated shapes seem likely (Phillipps & Driver 1995). Ellis et al. (1995) argue that the slope of the local luminosity function remains flat down to at least $M = -16$ and suggest that surface brightness dependent effects must act to preserve the flat shape of $\Phi(M)$. This occurs in the case of no correlation between surface brightness and luminosity (model A of Ferguson & McGaugh 1995), but I find this much less likely than the lack of correlation between surface brightness and size indicated directly by the data. This *must* have an effect on the faint end slope, though the upturn need not occur immediately faintwards of $L^*$.

## 6.2 Faint Blue Galaxies

The large excess of blue galaxies in faint number counts (e.g., Tyson 1988) led to many extreme models for galaxy evolution (e.g., Broadhurst et al. 1992). I argued (McGaugh 1994) that the correspondence of physical properties between populations of faint and low surface brightness galaxies suggested a common nature, removing the need for enormous amounts of evolution at low redshift. However, the implication that there is a one to one correspondence between local LSB galaxies and distant faint blue galaxies is incorrect – number counts do not work this way in the cosmological context. One can only discuss populations statistically, with the important effects being that of differential luminosity density measurements between surveys and potential underestimation of the slope of the luminosity function (Ferguson & McGaugh 1995).

There is now unambiguous evidence of evolution in the galaxy luminosity function (Ellis et al. 1995; Lilly et al. 1995a: CFRS VI). The CFRS is particularly persuasive, since it is deep and uniform and Lilly et al. (1995b: CFRS I) have thoroughly considered the relevant effects, including those discussed here. (Note that the CFRS does not imply a low density of LSB galaxies as might be inferred from Fig. 9 of CFRS I. Since it is a flux selected sample, one expects LSB galaxies to constitute a small fraction of the total sample even though easily detected: recall Fig. 3.) With a measurement isophote of $\mu_\ell = 28\ I_{AB}$ mag arcsec$^{-2}$, the CFRS detects nearly all the light for galaxies well into the VLSB regime, sufficiently far that very little residual luminosity is missed (Fig. 7). That the CFRS indicates a steep faint end slope for blue objects is thus no surprise. However, it primarily constrains the population at $z > 0.3$ where the observed amount of evolution is reasonable, if still large. The situation locally continues to pose the most serious problems. An upward revision of the local $\phi^*$ as suggested by several lines of evidence (see Ellis et al. 1995) cures many ills, but still provides no explanation for the steep galaxy counts at bright magnitudes (Maddox et al. 1990b) which imply a large amount of evolution very recently ($z < 0.1$), a significantly bluer galaxy population than usually assumed (Koo, Gronwall, & Bruzual 1993; see also Ferguson & McGaugh 1995; Babul & Ferguson 1995), or systematic errors (Metcalfe et al. 1995).

The effects discussed in this paper contribute to the solution of these problems, with an upward revision of both $\phi^*$ and $\alpha$ seeming likely. However, they are not of sufficient magnitude to be the sole solution. A factor of 2 in $\phi^*$ is required, at the upper limit of what can be attributed to surface brightness measurement effects. It is unclear what the faint end slope actually is. It seems plausible that red, early type galaxies have a flat luminosity function while a steeper one would be appropriate to the late types which make up the excess in the counts (Marzke et al. 1994; Glazebrook et al. 1995; Driver et al. 1995), especially as late types tend to be ISB and LSB galaxies (McGaugh et al. 1995b). If so, the type dependent redshift distribution will peak at lower redshifts for later types, or possibly even have a bimodal distribution if the classification scheme also includes strange things at high redshift in the late type bin. Since morphology is not quantitative, one really needs detailed knowledge of the trivariate distribution $\Phi(\mu_0, h, colour)$ in order to interpret evolutionary or cosmological effects. If the slope of this distribution becomes steep for blue, low surface brightness galaxies, it would cause an 'excess' which is not well constrained by local surveys (cf. Phillipps & Driver 1995).

### 6.3 Ly$\alpha$ Absorption Systems

Another quantity which requires the bivariate distribution is the cross section of galaxies as absorbers along the sight of QSOs. Indeed, one really needs to know the bivariate distribution of the gas discs. However, Fig. 8 provides a more direct way of estimating how much absorption is caused by discs of various surface brightnesses, as it gives an indication of how much gas mass is contained in each population. Though LSB galaxies are gas rich as individuals, the total H I density does decline with surface brightness.

Damped Ly$\alpha$ and Mg II absorption systems contain most of the mass seen in absorption. From Fig. 8, it is clear that most of these would be HSB and ISB systems readily visible optically. A significant but small fraction should be LSB galaxies, but these too should be fairly easily visible optically and so one does not expect the absorbers to remain unidentified when closely examined. Indeed, one example of an LSB galaxy being the absorber has recently been noted (Steidel et al. 1994).

There are low column density Ly$\alpha$ absorbers without optical counterparts (Morris et al. 1993). These systems do not contain much mass, but do exist in profusion. By extrapolation of Figures 5 and 8 into the ELSB regime, one does expect there to exist some such invisible galaxies, though they should contain very little mass. One might expect these ELSB galaxies to be weakly clustered (Mo et al. 1994), consistent with observations of the Ly$\alpha$ clouds (Mo & Morris 1994). However, the extrapolations involved are huge so the uncertainties become enormous. At most I would say that it plausible that the low column density absorption systems are related to a population of ELSB galaxies. It is just as plausible that they are bubble-like structures unrelated to individual galaxies.

## 6.4 Passband Biases: A Cautionary Tale

The data discussed here are all selected from blue sensitive photographic plates. The known examples of very large, massive, VLSB galaxies similar to Malin 1 tend to be rather more red than the normal sized LSB population (Sprayberry et al. 1995a). This raises the concern that substantial numbers of massive red LSB galaxies may still be missed.

Indeed, despite the large amounts of effort that have gone into galaxy surveys, the situation even in the blue remains fairly abysmal. In the red, the appropriate data do not exist at all. In order to examine possible passband biases, consider the case of selection on blue sensitive plates for galaxies of the same bolometric surface brightness but a range of colours. In analogy with Fig. 2, imagine that equal numbers of disc galaxies exist at each bolometric surface brightness. The redder discs will, for the same $(\mu_0^{bol}, h)$, appear smaller on blue sensitive plates than blue discs. Hence fewer red discs will be selected at any given $\theta_\ell$.

This is illustrated in Fig. 9, which shows the relative number of galaxies one would expected to detect as a function of $B - I$ colour for representative combinations of bolometric central surface brightness and limiting $B$ isophote. The results depend sensitively on the passband and the exact shape of the spectral energy distribution, especially as $\mu_0^{bol}$ approaches $\mu_\ell$. There is no way to know what spectral energy distributions are really appropriate for the putative population of red LSB galaxies. These could be faded star bursts (and indeed, one would expect significant numbers of such remnants from the high number of known short duty cycle star burst dwarfs) or something entirely different. For the sake of illustration, the spectral energy distributions from the models of Buzzoni (1991) are utilised.

Even for HSB galaxies selected at a fairly deep $\mu_\ell$, the apparent number of galaxies declines steadily as their colours become redder. The effect becomes more severe as either $\mu_0^{bol}$ becomes fainter or $\mu_\ell$ brighter. Galaxies begin to disappear entirely for only modestly red colours ($B - I \approx 2$) for galaxies with $\mu_0^{bol} \gtrsim 23$, especially at the brighter limiting isophotes typical of large surveys. This is several magnitudes brighter than Malin 1, so it is worth emphasising that we would still be unaware of this enormous galaxy had it lacked a prominent bulge component. All of the *known* examples of galaxies similar to Malin 1 also have prominent bulges even though most LSB galaxies have little or no bulge.

The distribution of colours for LSB galaxies in the Schombert et al. (1992) catalog roughly follows the appropriate line in Fig. 9. This implies a density of red LSB galaxies as large as that of blue LSB galaxies (i.e., a roughly flat distribution with colour). The data are too limited to say more because the volume probed for red LSB galaxies by $B$-band surveys is extremely small. Considering the brightness of the night sky at redder wavelengths, and that contrast for blue LSB galaxies is already a problem in $B$, it could well be that only the most prominent, observationally accessible component of the galaxy population has been identified. An entire universe full of dim galaxies may remain to be discovered.

## 7 CONCLUSIONS

I have quantified the relative number, luminosity, and mass density of disc galaxies as a function of central surface brightness. These relations can be conveniently expressed analytically as

$$\log[\phi(\mu_0)] = m(\mu_0 - \mu_0^*)$$

for the number density,

$$\log[J(\mu_0)] = (m - 0.4)(\mu_0 - \mu_0^*)$$

for the luminosity density, and

$$\log[\rho(\mu_0)] = (m - 0.2)(\mu_0 - \mu_0^*)$$

for the mass density. This last refers to both the dynamical mass enclosed within the optical radius, and to the H I gas mass. See equations (25), (28), (31), and (32) for further details. The values of the parameters $\mu_0^*$ and $m$ obtained from extant data are

$$\mu_0^* = 21.5 \ B_J \text{ mag arcsec}^{-2},$$

$$m = 2.6 \qquad \text{for} \quad \mu_0 < \mu_0^*, \text{ and}$$

$$-0.3 \leq m \leq 0.1 \qquad \text{for} \quad \mu_0 > \mu_0^*.$$

These relations are consistent with all relevant data, though substantial uncertainties remain. They represent a major shift from the paradigm of the Freeman law which suggests that spiral galaxies all have nearly the same central surface brightness. The value of $\mu_0^*$ corresponds to Freeman's value, but a faint end slope of $m \approx 0$ indicates roughly equal numbers of galaxies at each central surface brightness fainter than $\mu_0^*$. This confirms the intuition of Disney (1976) that Freeman's law is a selection effect, but one which works more in the fashion described by Allen & Shu (1979).

It is very striking that the surface brightness distribution is nearly flat. It is unclear what significance, if any, should be attached to the particular value $m = 0$. The sharp cutoff brightwards of $\mu_0^*$ seems analogous to the turndown in the luminosity function brightwards of $L^*$, but the significance of the particular value of $\mu_0^*$ is also unclear. What is clear is that these numbers are fundamental to understanding spiral galaxies.

## ACKNOWLEDGEMENTS

This paper was written over the course of two years while I was supported by a PPARC PDRA, for which I am grateful. During that time I have aspired to keep the manuscript current with the many relevant developments, and apologize for any I may have inadvertantly overlooked. I would like to thank the many people, too numerous to list, who have tolerated my pestering them about these issues and who responded with many lively and often incredulous conversations. Deserving of special mention are Greg Aldering, Harry Ferguson, Roelof de Jong, David Schade, Bob Abraham, and Joseph Lehár.

# REFERENCES

Allen, R. J., & Shu, F. H. 1979, ApJ, 227, 67

Antonuccio-Delogu, V. 1995, Astrophys. Letters and Comm., in press

Babul, A. & Ferguson, H. C. 1995, ApJ, in press

Boroson, T. 1981, ApJS, 46, 177

Bosma, A. & Freeman, K. C. 1993, AJ, 106, 1394

Bothun, G. D., Impey, C. D., & Malin, D. F. 1991, ApJ, 376, 404

Bothun, G. D., Impey, C. D., Malin, D. F., & Mould, J. R. 1987, AJ, 94, 23

Bothun, G. D., Schombert, J. M., Impey, C. D., & Schneider, S. E. 1990, ApJ, 360, 427

Bothun, G. D., Schombert, J. M., Impey, C. D., Sprayberry, D., & McGaugh, S. S. 1993, AJ, 106, 530

Briggs, F. H. 1990, AJ, 100, 999

Broadhust, T. J., Ellis, R. S., & Glazebrook, K., 1992, Nature, 355, 55

Buzzoni, A. 1989, ApJS, 71, 817

Cornell, M., Aaronson, M., Bothun, G. D., & Mould, J. 1987, ApJS, 64, 507

Corwin, H. G., de Vaucoulers, A., & de Vaucoulers, G. 1980, AJ, 85, 1027

Dalcanton, J. J. 1995, Ph.D. thesis, Princeton University

Dalcanton, J. J., Spergal, D. N. & Summers, F. J. 1995, preprint

Davies, J. I. 1990, MNRAS, 244, 8

Davies, J. I., Phillipps, S., Cawson, M. G. M., Disney, M. J., & Kibblewhite, E. J. 1988, MNRAS, 232, 239

Davies, J. I., Phillipps, S., Disney, M. J 1988, MNRAS, 231, 69P

Davies, J. I., Phillipps, S., Disney, M. J., Boyce, P., & Evans, Rh. 1994, MNRAS, 268, 984.

de Blok, W. J. G., van der Hulst, J. M., & Bothun, G. D. 1995a, MNRAS, 274, 235

de Blok, W. J. G., McGaugh, S. S., & van der Hulst 1995b, submitted

de Jong, R. S. 1995, Ph.D. thesis, University of Groningen

de Jong, R. S., & van der Kruit, P. C. 1994, A&AS, 106, 451

de Vaucoulers, G. 1948, Ann. d'Astrophys., 11, 247

de Vaucoulers, G. 1959, in Handbuch der Physik, vol. 53, Astrophysics IV: Stellar Systems, ed. S. Flügge (Berlin: Springer-Verlag), 275

de Vaucoulers, G. 1974, in IAU Symposium 58, Formation and Dynamics of Galaxies, ed. J. R. Shakeshaft (Dordrecht: Reidel), 1

Disney, M. J. 1976, Nature, 263, 573

Disney, M. J., & Phillipps, S. 1983, MNRAS, 205, 1253

Driver, S. P., Windhorst, R. A., & Griffiths, R. E. 1995, ApJ, 453, 48

Ellis, R. S., Colless, M., Broadhurst, T., Heyl, J., & Glazebrook, K. 1995, MNRAS, in press

Ferguson, H. C., & McGaugh, S. S. 1995, ApJ, 440, 470

Frenk, C. S., White, S. D. M., Davis, M., & Efstathiou, G. 1988, ApJ, 327, 507

Fisher, J. R., & Tully, R. B. 1981, ApJ, 243, L23

Freeman, K. C. 1970, ApJ, 160, 811

Giovanelli, R., & Haynes, M. P. 1989, ApJ, 346, L5

Giovanelli, R., Haynes, M. P., Salzer, J. J., Wegner, G., Da Costa, L. N., & Freudling, W. 1995, AJ, 110, 1059

Glazebrook, K., Ellis, R. S., Santiago, B., & Griffiths, R. 1995, MNRAS, 275, L19

Hudson, M. J., & Lynden-Bell, D. 1991, MNRAS, 252, 219

Impey, C. D., & Bothun, G. D. 1989, ApJ, 341, 89

Impey, C. D., Bothun, G. D., & Malin, D. F., Staveley-Smith, L. 1990, ApJ, 351, L33

Irwin, M. J., Davies, J. I., Disney, M. J., & Phillipps, S. 1990, MNRAS, 245, 289

Kent, S. M. 1985, ApJS, 59, 115

Kerr, F. J., & Henning, P. A. 1987, ApJ, 320, L99

Knezek, P. 1993, Ph.D. thesis, University of Massachusetts

Koo, D. C. & Kron, R. G. 1992, ARA&A, 30, 613

Koo, D. C., Gronwall, C., & Bruzual, G. A. 1993, ApJ, 415, L21

Kormendy, J. 1977, ApJ, 217, 406

Lauberts, A. & Valentijn, E. A. 1989, The Surface Photometry Catalog of the ESO-Uppsala Galaxies (Munich: ESO)

Lilly, S.J., Le Fevre, O., Crampton. D., Hammer, F.,Tresse, L., 1995b, ApJ, in press (CFRS I)

Lilly, S.J., Tresse, L., Hammer, F., Le Fevre, O., Crampton, D., 1995a, ApJ, in press (CFRS VI)

Loveday, J., Peterson, B. A., Efstathiou, G., & Maddox, S. J. 1992, ApJ, 390, 338

Maddox, S. J, Sutherland, W. J., Efstathiou, G., & Loveday, J. 1990a, MNRAS, 243, 692

Maddox, S. J, Sutherland, W. J., Efstathiou, G., Loveday, J., & Peterson, B. A. 1990b, MNRAS, 1P

Marzke, R. O., Geller, M. J., Huchra, J. P., & Corwin, H. G. 1994, AJ, 108, 437

McGaugh, S. S. 1992, Ph.D. thesis, University of Michigan

McGaugh, S. S. 1993, BAAS, 25, 1384

McGaugh, S. S. 1994, Nature, 367, 538

McGaugh, S. S. 1995, in New Light on Galaxy Evolution, IAU Symp. No. 171, ed. R. Bender & R. Davies (Kluwer), in press

McGaugh, S. S. & Bothun, G. D. 1994, AJ, 107, 530

McGaugh, S. S., Bothun, G. D., & Schombert, J. M. 1995a, AJ, 110, 573

McGaugh, S. S., Schombert, J. M., & Bothun, G. D. 1995b, AJ, 109, 2019

McGaugh, S. S., de Blok, W. J. G., van der Hulst, J. M., & Zwaan, M. A. 1995, submitted

McMahon, R. G., Irwin, M. J., Gioovanelli, R., Haynes, M. P., & Wolfe, A. M. 1990, ApJ, 359, 302

Metcalfe, N., Shanks, T., Fong, R., & Roche, N. 1995, MNRAS, 273, 257

Mo, H. J., McGaugh, S. S., & Bothun, G. D. 1994, MNRAS, 267, 129

Mo, H. J., & Morris, S. L. 1994, MNRAS, 269, 52

Morris, S. L., Weymann, R. J., Dressler, A., McCarthy, P. J., Smith, B. A., Terrile, R. J., Giovanelli, R., & Irwin, M. 1993, ApJ, 419, 524

Nilson, P. 1973, Uppsala General Catalog of Galaxies, (Uppsala, Sweden: Societatis Scientiarum Upsaliensis) (UGC)

Paturel, G. 1975, A&A, 40, 133

Peletier, R. F., & Willner, S. P. 1992, AJ, 103, 1761

Phillipps, S., & Disney, M. J. 1983, MNRAS, 203, 55

Phillipps, S., & Disney, M. J. 1986, MNRAS, 221, 1039

Phillipps, S., Disney, M. J., Kibblewhite, E. J., & Cawson, M. G. M. 1987, MNRAS, 229, 505

Phillipps, S., & Driver, S. 1995, MNRAS, in press.

Rao, S., & Briggs, F. 1993, ApJ, 419, 515

Romanishin, W., Strom, K. M., & Strom, S. E. 1983, ApJS, 53, 105

Rönnback, J. 1993, Ph.D. thesis, Uppsala University

Roukema, B. F., & Peterson, B. A. 1995, A&AS, 109, 511

Salpeter, E. E., & Hoffman, G. L. 1995, in preparation

Salzer, J. J., Alighieri, S. D. S., Matteucci, F., Giovanelli, R., & Haynes, M. P. 1991, AJ, 101, 1258

Schade, D. J., & Ferguson, H. C. 1994, MNRAS, 267, 889

Schechter, P. 1976, ApJ, 203, 297

Schombert, J. M., & Bothun, G. D. 1987, AJ, 93, 60

Schombert, J. M., & Bothun, G. D. 1988, AJ, 95, 1389

Schombert, J. M., Bothun, G. D., Schneider, S. E., & McGaugh, S. S. 1992, AJ, 103, 1107

Schwartzenberg, J. M., Phillipps, S., Smith, R. M., Couch, W. J., & Boyle, B. J. 1995, MNRAS, 275, 121

Schweizer, F. 1976, ApJS, 31, 313

Sérsic, J.-L. 1969, Atlas de galaxias australes (Observatorio Astronomica, Córdoba)

Sodré, L., & Lahav, O. 1993, MNRAS, 260, 285

Sprayberry, D. 1994, Ph.D. thesis, University of Arizona

Sprayberry, D., Bernstein, G. M., Impey, C. D., & Bothun, G. D. 1995b, ApJ, 438, 72

Sprayberry, D., Impey, C. D., Bothun, G. D., & Irwin, M. 1995a, AJ, 109, 558

Sprayberry, D., Impey, C. D., Irwin, M., McMahon, R. G., & Bothun, G. D. 1993, ApJ, 417, 114

Steidel, C. C., Pettini, M., Dickinson, M., & Persson, S. E. 1994, AJ, 108, 2046

Szomoru, A., Guhathakurta, P., van Gorkom, J. H., Knapen, J. H., Weinberg, D. H., & Fruchter, A. S. 1995, AJ, in press

Thuan, T. X., & Seitzer, P. O. 1979, ApJ, 231, 680

Tyson, A. J. 1988, AJ, 96, 1

van der Kruit, P. C. 1987, A&A, 173, 59

Weinberg, D. H., Szomoru, A., Guhathakurta, P., & van Gorkom, J. H. 1991, ApJ, 372, L13

Zwaan, M. A., van der Hulst, J. M., de Blok, W. J. G., & McGaugh, S. S. 1995, MNRAS, 273, L35

Zwicky, F. 1957, Morphological Astronomy (Berlin: Springer-Verlag)

# FIGURE CAPTIONS

**Figure 1.** Idealised galaxy luminosity profiles for four exponential discs. The galaxies are characterised by their size $h$ and central surface brightness $\mu_0$. The HSB galaxies are Freeman discs with $\mu_0 = 21.65$ $B$ mag arcsec$^{-2}$ and the LSB disks have $\mu_0 = 23.15$. The scale lengths are chosen to illustrate the variation of isophotal quantities at fixed size and luminosity for different surface brightnesses. The HSB disc with $h = 3$ kpc has an integrated luminosity of $8 \times 10^9$ $L_\odot$, roughly $L^*$ for $H_0 = 100$ km s$^{-1}$Mpc$^{-1}$. The LSB disc with $h = 6$ kpc has this same luminosity, while the LSB disc with $h = 3$ kpc has the same physical size but only one quarter the luminosity. The small $h = 1.5$ kpc HSB disc also has $L = \frac{1}{4}L^* = 2 \times 10^9$ $L_\odot$. Note that when measured at a particular isophotal level (e.g., the horizontal lines at $\mu_\ell = 24$ and $\mu_\ell = 26$), an LSB galaxy will always appear smaller than an HSB galaxy of the same size, and fainter than one of the same luminosity.

**Figure 2.** The relative number of galaxies observed as a function of central surface brightness for diameter limited catalogs. The solid line shows the assumed intrinsic distribution with equal numbers at every $\mu_0$ faintwards of the value $\mu_0^* \approx 21.5$ $B_J$ mag arcsec$^{-2}$. Brighter than $\mu_0^*$ the intrinsic distribution is assumed to have a sharp exponential cut off to be consistent with observations. The constant density assumed on the faint side illustrates the effects of volume sampling for diameter limited catalogs selected at different isophotal levels $\mu_\ell$. The resultant apparent distributions (dashed lines) are shown (normalised to $N_{obs} = 100$ at $\mu_0^*$) for several representative $\mu_\ell$ (labeled). These are typical of shallow plate material ($\mu_\ell = 24$), deep plate material ($\mu_\ell = 26$), and very deep CCD images ($\mu_\ell = 28$). The apparent distribution is very sharply peaked at the value $\mu_0^*$ for all values of $\mu_\ell$, giving the appearance that all galaxies have $\mu_0 \approx \mu_0^*$ even if equal numbers of galaxies exist at every $\mu_0 > \mu_0^*$.

**Figure 3.** As per Fig. 2, but for catalogs limited by isophotal magnitude. The apparent distribution of $\mu_0$ is even sharper than in the diameter limited case because $N_{obs} \propto 10^{-0.6(\mu_0 - \mu_0^*)}$ rather than $(\mu_\ell - \mu_0)^3$.

**Figure 4.** The distribution of diameter ratios $\Gamma$, which is related to the surface brightness $\mu_0$. The histogram is the SRC/ESO data of Bosma & Freeman (1993). The heavy solid line shows the expected apparent distribution for a flat intrinsic distribution. This is *very* sharply peaked because of the way $\Gamma$ is defined. The dotted line gives the expected distribution after convolution with errors, indicating that these data suggest an intrinsically flat distribution.

**Figure 5.** The surface brightness distribution derived from the data of Davies (1990). Circles are data selected by diameter, while triangles are magnitude limited data. Error bars are from counting statistics. The lines are least squares fits to the data giving equal weight to points from diameter and magnitude limited samples. The distribution declines slowly faintwards of the Freeman value, indicating a large space density of low surface brightness galaxies. It cuts off sharply brightwards of $\mu_0^*$ in analogy with the turndown of the luminosity function at $L^*$.

**Figure 6.** The surface brightness distribution after maximally extreme corrections for the effects of cosmological dimming. Triangles: the flux selected data shifted by the mean correction. Dashed line: redistribution assuming all the dimmest galaxies are the highest redshift ones. Dot dashed line: redistribution given the opposite assumption (see text). These lines give the maximum envelope to which cosmological dimming can modulate the surface brightness distribution derived from the data of Davies (1990). This can not be an important effect, as it would cause inconsistency with the data of Schombert et al. (1992).

**Figure 7.** The luminosity density produced by disc galaxies as a function of surface brightness. The luminosity density peaks at the Freeman value, and declines steadily towards fainter surface brightnesses. Substantially more luminosity is produced by discs which are fainter than $\mu_0^*$ than is produced by those which are brighter. In addition to the data of Davies (1990), the data of Schombert et al. (1992) are plotted as the solid square. For this, the solid line shows the error bar due to counting statistics alone, while the dashed line shows the additional estimate of systematic error.

**Figure 8.** The mass density contained in disc galaxies as a function of surface brightness. Points as per Fig. 7. A significant amount of mass is contained in the low surface brightness tail of the distribution in excess of that in Fig. 7 because the mass to light ratio is observed to increase as the surface brightness decreases. Note that LSB discs with $\mu_0 = 23$ contain $\gtrsim 1/3$ the amount of mass that is in galaxies with $\mu_0 = \mu_0^* = 21.5$.

**Figure 9.** The apparent number of galaxies of the same bolometric surface brightness that will be observed as a function of color. Galaxies are assumed to exist with an intrinsically flat distribution (equal number density at each $\mu_0^{bol}$ and color), and are selected by diameter on blue sensitive plates at various isophotal levels $\mu_\ell^B$. The plot is normalised to $N_{obs} = 100$ for $B - I = 1.3$, the typical color of a $B$ selected LSB galaxy. Though $B$-band surveys provide the best contrast for LSB galaxies since the sky is reasonably dark there, they are quite insensitive to red LSB galaxies. The exact sensitivity depends on the limiting isophote and the shape of the spectral energy distribution in a complicated way.

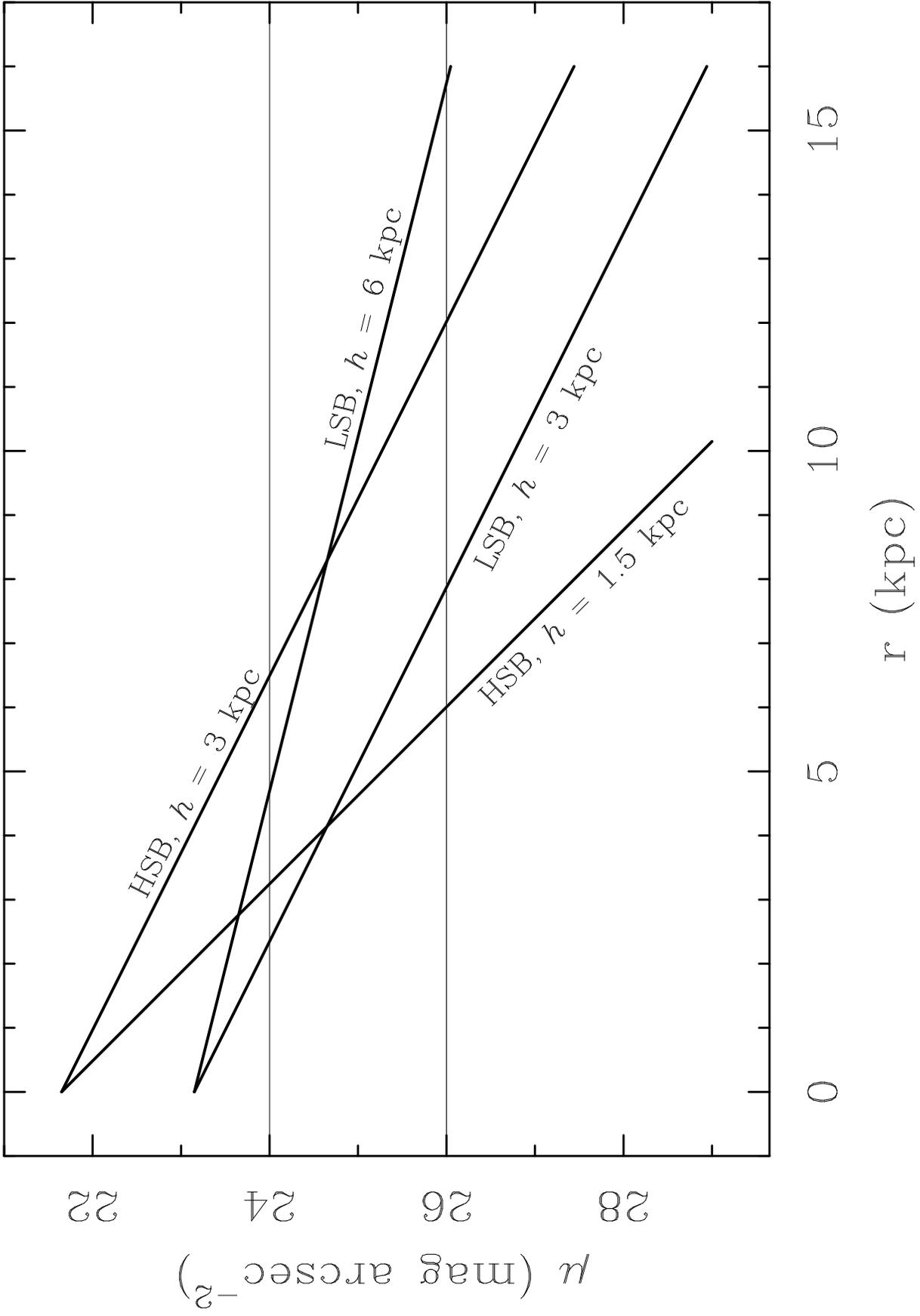

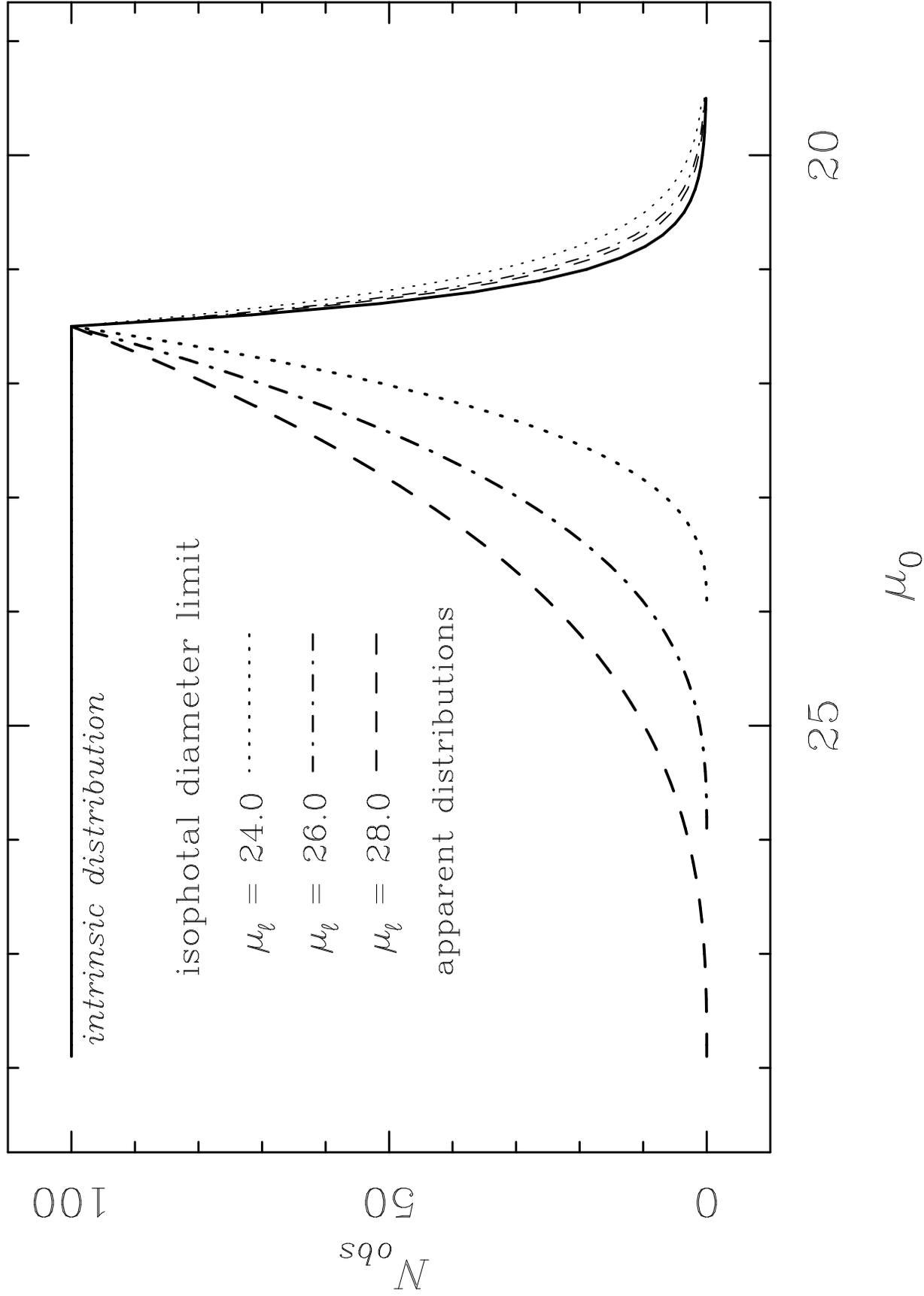

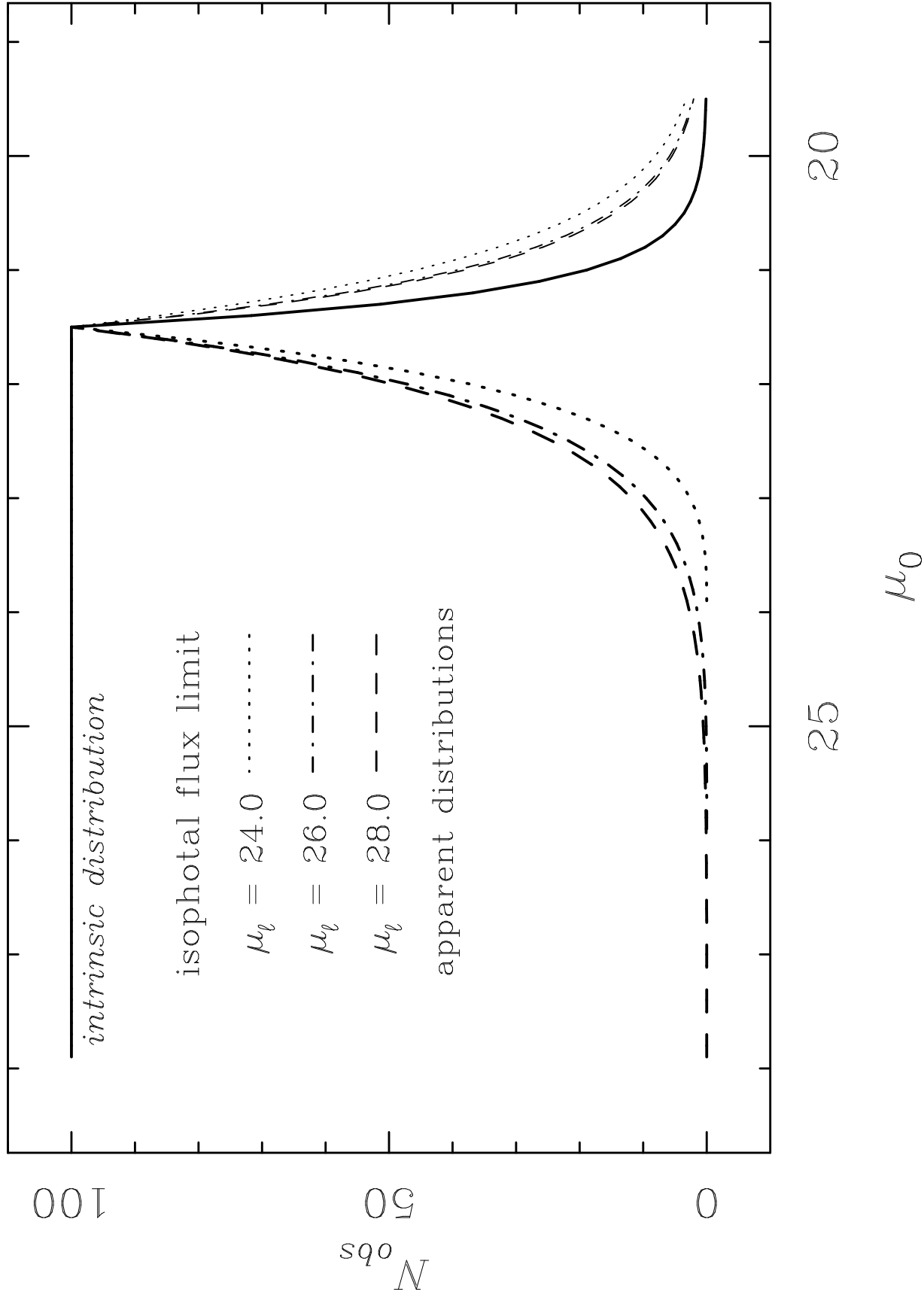

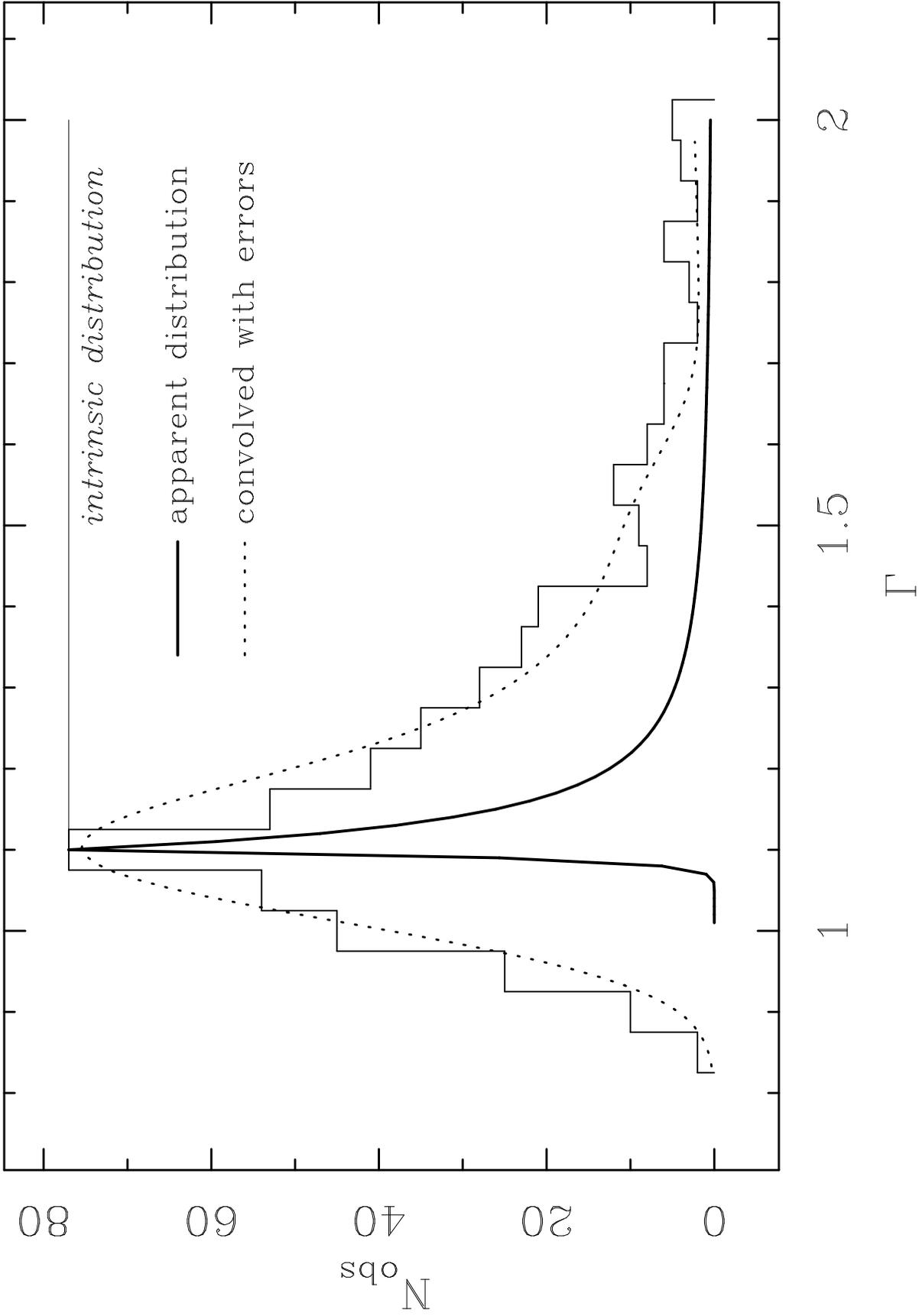

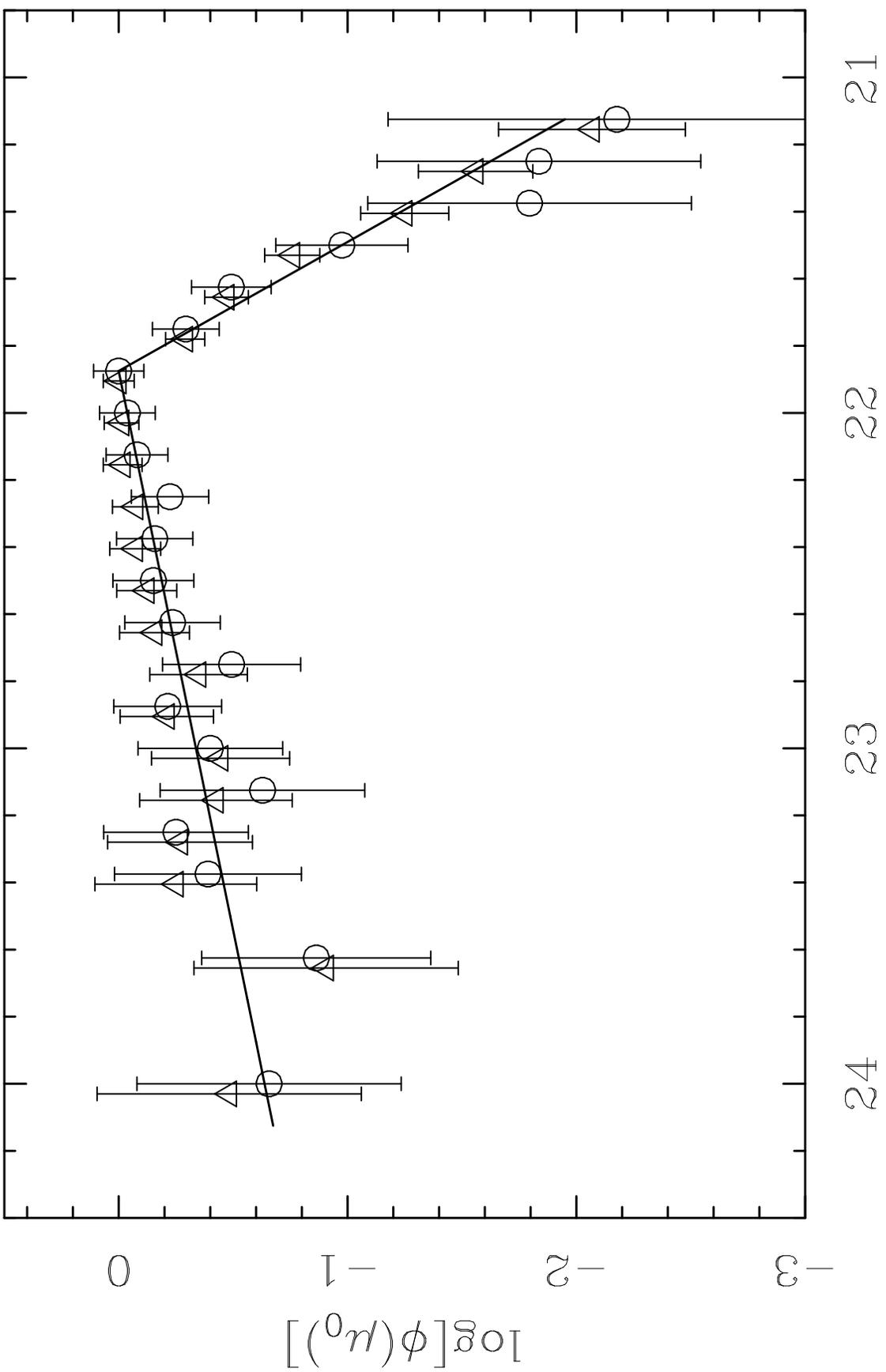

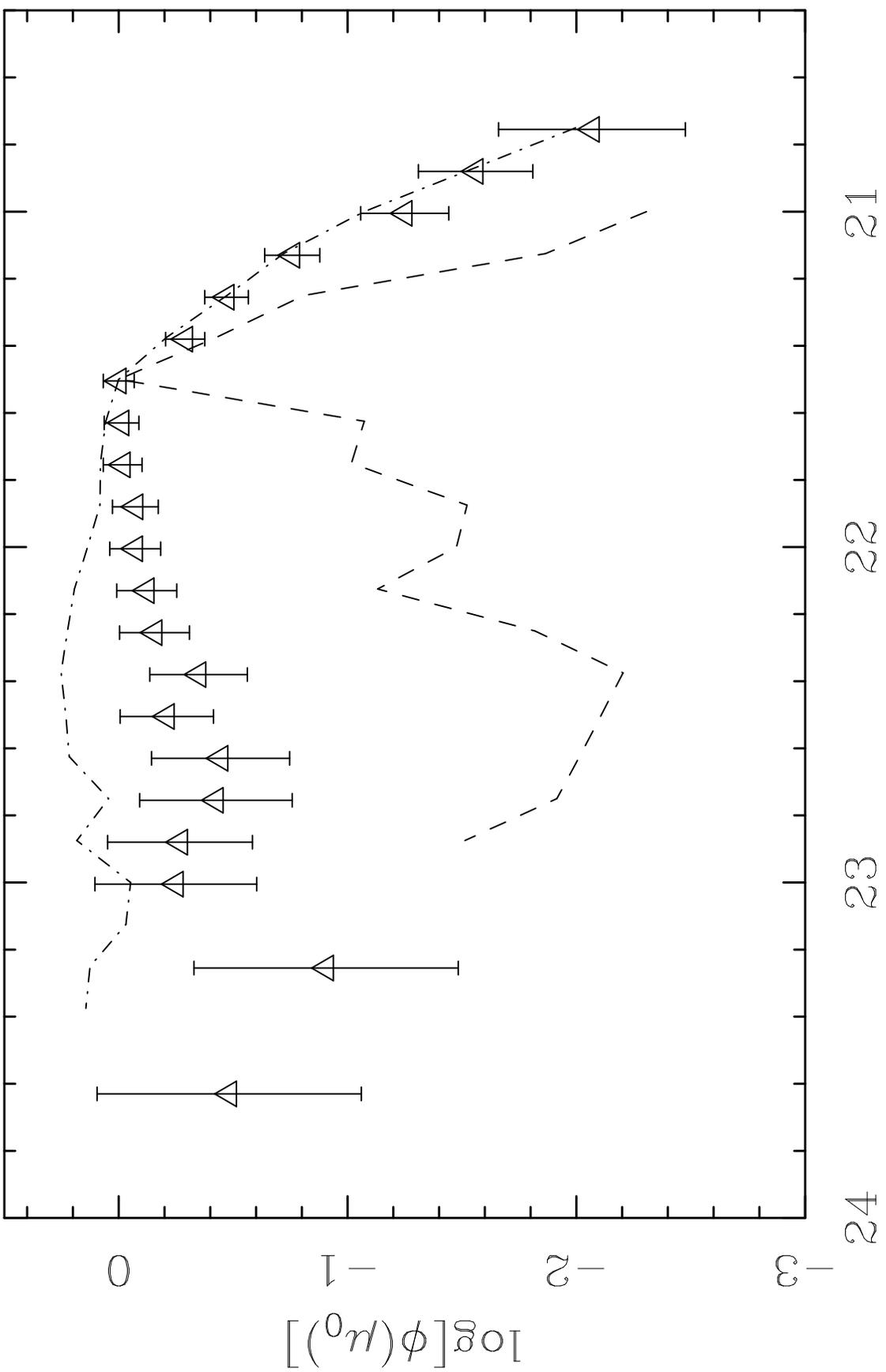

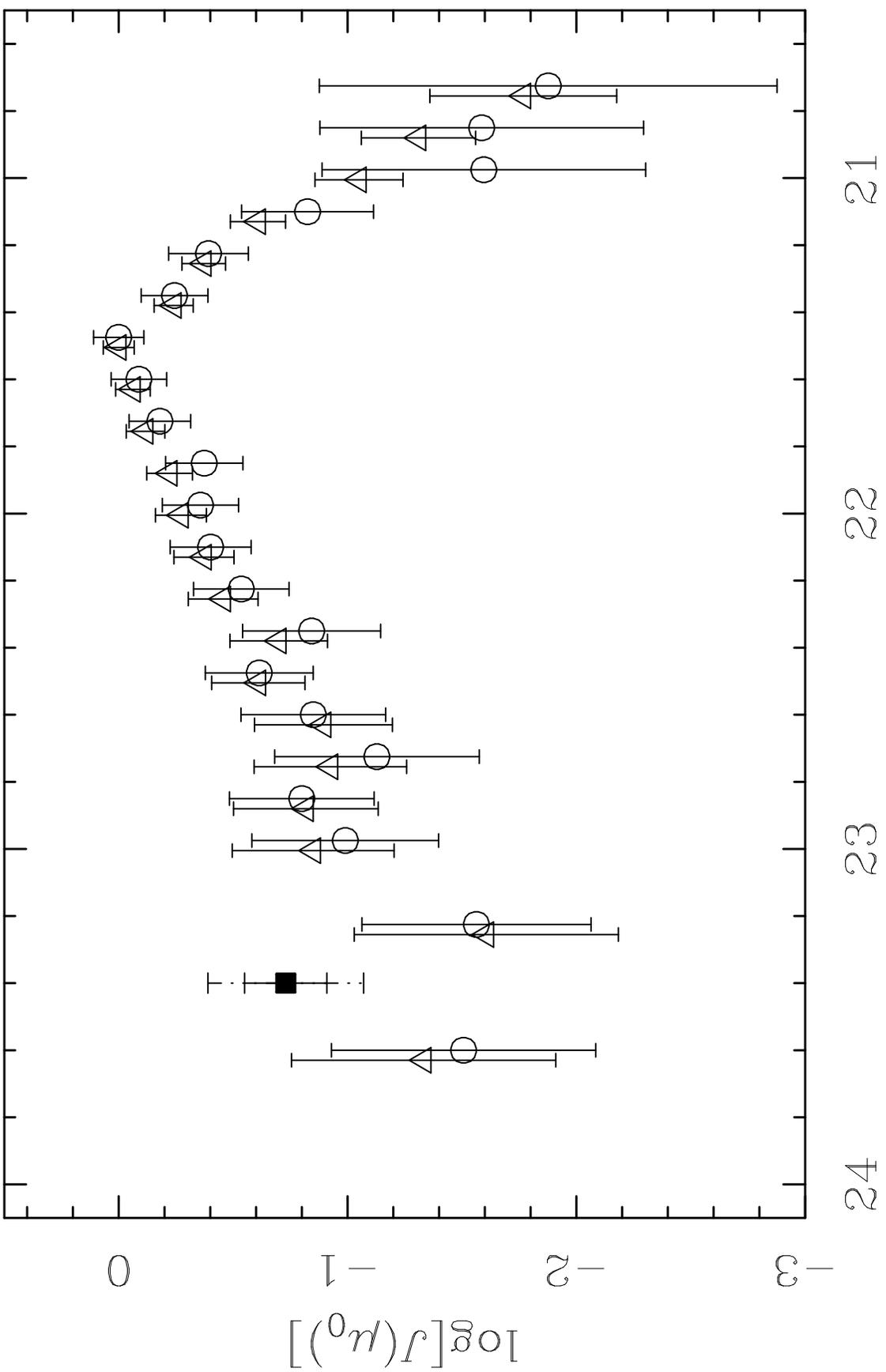

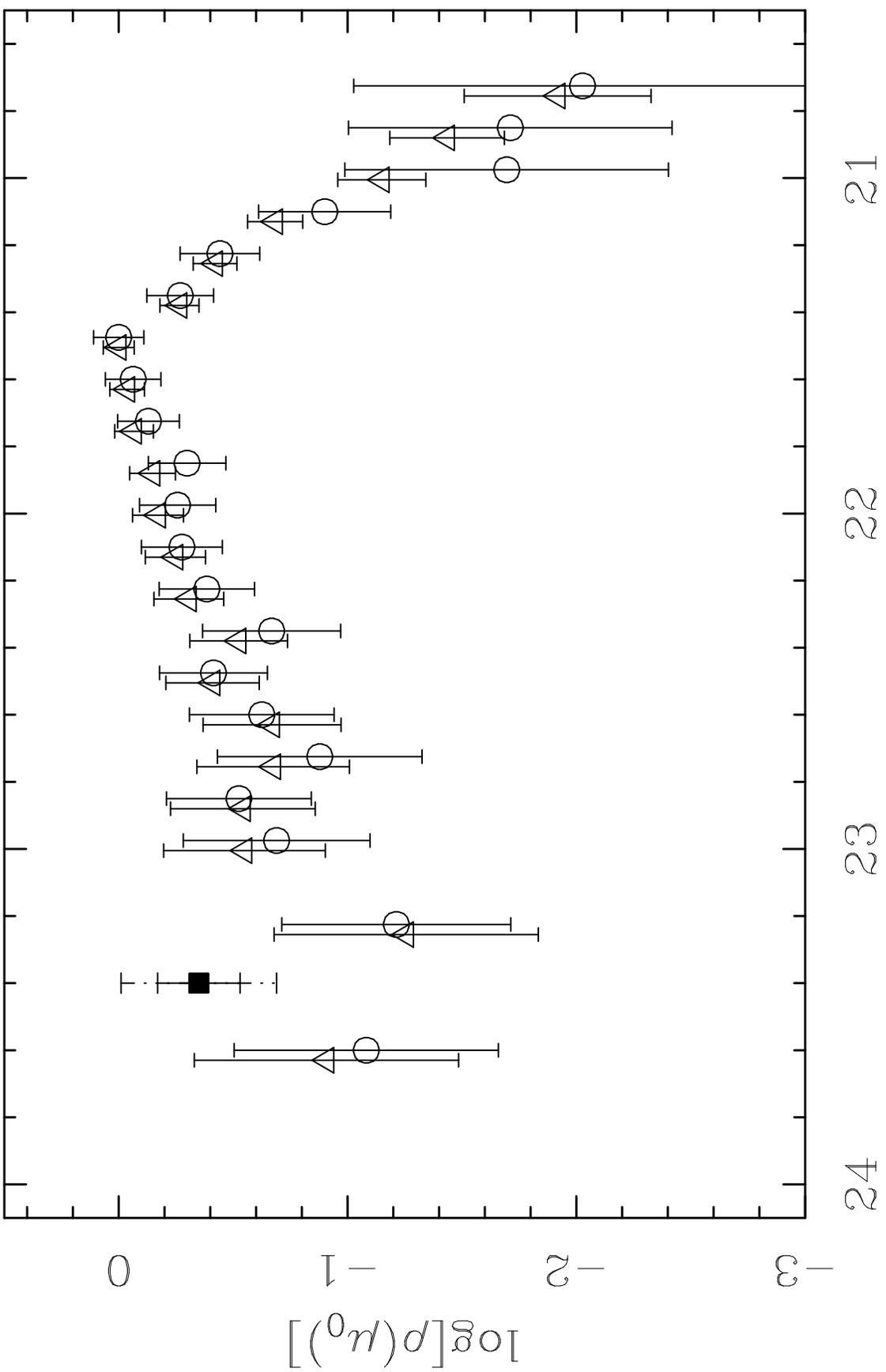

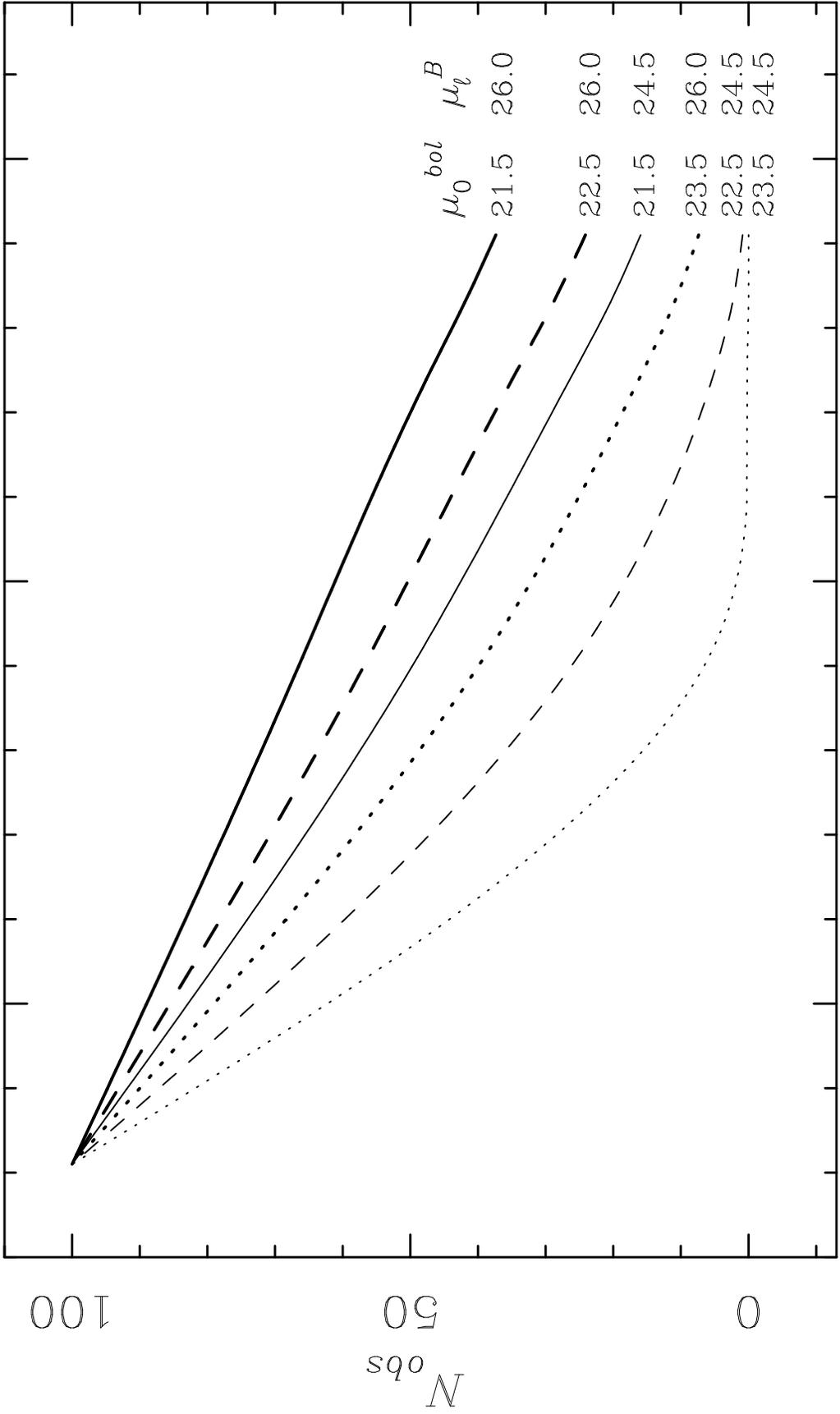

**Table 1.** Nomenclature

| Name | $\mu_0$ | $\Sigma_0$ |
|---|---|---|
| | $B$ mag arcsec$^{-2}$ | $L_\odot$pc$^{-2}$ |
| VHSB | $< 21.25$ | $> 200$ |
| HSB[a] | $21.25 - 22.0$ | $100 - 200$ |
| ISB | $22.0 - 22.75$ | $50 - 100$ |
| LSB | $> 22.75$[b] | $< 50$ |
| VLSB | $24.5 - 27.0$ | $1 - 10$ |
| ELSB[c] | $> 27.0$ | $< 1$ |

[a]Satisfy Freeman's Law
[b]Brightness of darkest night sky
[c]Practically invisible

**Table 2.** Variations

| $\mu_\ell^U$ | $\mu_\ell^H$ | $\mu_0^L$ | $h_H/h_L$ | $log(n_L/n_H)$ |
|---|---|---|---|---|
| 25.3 | 26.5 | 23.4 | 1 | $-0.03$ |
| $<24.3$ | 26.5 | 23.4 | 1 | $-0.24$ |
| 25.3 | 26.0 | 23.4 | 1 | $-0.16$ |
| 25.3 | 26.5 | 23.0 | 1 | $-0.17$ |
| 25.3 | 26.5 | 24.0 | 1 | $0.24$ |
| 25.3 | 26.5 | 23.4 | 2 | $0.87$ |
| 25.3 | 26.5 | 23.4 | 0.5 | $-0.93$ |
| 24.3 | 26.0 | 23.0 | 1 | $-0.53$ |